\documentclass[twocolumn,prb,reprint,superscriptaddress,floatfix,amsmath,amssymb,aps]{revtex4-2}
\pdfoutput=1
\usepackage[normalem]{ulem}
\usepackage[utf8]{inputenc}
\usepackage[caption=false,subrefformat=parens,labelformat=parens]{subfig}
\usepackage{graphicx}
\usepackage{float}
\usepackage{amsmath}
\usepackage{dcolumn}
\usepackage{bm}
\usepackage{xcolor}
\usepackage{xfrac}
\usepackage[colorlinks=true, allcolors=blue]{hyperref}
\usepackage{chemformula}
\usepackage{siunitx}

\def\a{\alpha}
\def\b{\beta}

\begin{document}
\title{Giant coercivity and enhanced intrinsic anomalous Hall effect at vanishing magnetization in a compensated kagome ferrimagnet}

\author{Jonathan M. DeStefano}
\affiliation{Department of Physics, University of Washington, Seattle WA, 98112, USA}
\author{Elliott Rosenberg}
\affiliation{Department of Physics, University of Washington, Seattle WA, 98112, USA}
\author{Guodong Ren}
\affiliation{Materials Sciences \& Engineering Department, University of Washington, Seattle WA, 98115, USA}
\author{Yongbin Lee}
\affiliation{Ames Laboratory, U.S. Department of Energy, Ames, Iowa 50011, USA}
\author{Zhenhua Ning}
\affiliation{Ames Laboratory, U.S. Department of Energy, Ames, Iowa 50011, USA}
\author{Olivia Peek}
\affiliation{Department of Physics, University of Washington, Seattle WA, 98112, USA}
\author{Kamal Harrison}
\affiliation{Department of Physics, University of Washington, Seattle WA, 98112, USA}
\affiliation{Department of Physics, University of Central Florida, Orlando FL, 32816, USA}
\affiliation{NanoScience Technology Center, University of Central Florida, Orlando FL, 32826, USA}
\author{Saiful I. Khondaker}
\affiliation{Department of Physics, University of Central Florida, Orlando FL, 32816, USA}
\affiliation{NanoScience Technology Center, University of Central Florida, Orlando FL, 32826, USA}
\affiliation{School of Electrical Engineering and Computer Science, University of Central Florida, Orlando, Florida 32826, USA}
\author{Liqin Ke}
\affiliation{Ames Laboratory, U.S. Department of Energy, Ames, Iowa 50011, USA}
\author{Igor I. Mazin}
\affiliation{Department of Physics and Astronomy and Quantum Science and Engineering Center, George Mason University, Fairfax, Virginia 22030, USA}
\author{Juan Carlos Idrobo}
\affiliation{Materials Sciences and Engineering Department, University of Washington, Seattle WA, 98115, USA}
\affiliation{Physical and Computational Sciences Directorate, Pacific Northwest National Laboratory, Richland WA, 99354, USA}
\author{Jiun-Haw Chu}
\email{jhchu@uw.edu}
\affiliation{Department of Physics, University of Washington, Seattle WA, 98112, USA}

\date{\today}
\begin{abstract}
Ferrimagnets that can be driven to magnetic compensation show promise for use in spintronics as they exhibit a finite anomalous Hall effect at zero magnetic field without having a significant magnetic moment. Compensated ferrimagnet spintronic devices with both a large anomalous Hall effect and a high coercivity would be simultaneously easy to read and difficult to erase. The kagome ferrimagnet \ch{TbMn6Sn6} has been reported to host a large intrinsic anomalous Hall effect. Here, we demonstrate that doping the Mn sites with Cr drives the system towards magnetic compensation. For nearly compensated compositions at low temperatures, giant coercive fields exceeding \SI{14}{T} are observed. Additionally, Cr doping significantly enhances the intrinsic anomalous Hall effect, which can be attributed to a shift in the Fermi level. Our results extend the range of unique magnetic states observed in kagome materials, demonstrating that chemical doping is an effective strategy to tune and realize these states.    
\end{abstract}

\maketitle

\section{Introduction}
Compensated ferrimagnets have recently emerged as a promising material platform for spintronics applications, combining advantageous properties of both ferromagnets and antiferromagnets~\cite{CompensatedFMreview, kimFerrimagneticSpintronics2022, zhouEfficientSpintronicsFully2021}.
In antiferromagnets, the net moment is zero by symmetry. This feature leads to faster switching times, reduced stray fields and resistance against external fields, but also leads to difficulty to read and write magnetic domains. A related class of Luttinger-compensated ferrimagnets~\cite{Mazin2022} also features exact compensation, albeit not by symmetry but by virtue of Luttinger theorem.
These can be insulators, or half-metals (metallic in one spin channel).
A more promising possibility are chemically compensated ferrimagnets, which are metallic in both spin channels. There, the net magnetization can be tuned to zero by controlling the chemical composition of their inequivalent spin sublattices. A great advantage here is that the net magnetic moment can be tunable by chemical composition or temperature, so that it is small, but not zero, thus allowing domain control by external fields. Another important advantage is that despite near-zero magnetization, compensated ferrimagnets allow for large (on the order of hundreds meV) exchange splitting. Consequently, they can exhibit finite anomalous Hall effect (AHE), providing a relatively straightforward reading mechanism. 

A well-known class of compensated ferrimagnets are the rare-earth transition-metal alloys~\cite{ReTMreview}. In these materials the magnetic moments of rare-earth and transition-metal ions are antiferromagnetically coupled, and the net magnetic moment  can be controlled by tuning their relative concentrations. The magnetization of the rare-earth and the transition-metal moments have different temperature dependencies. As a result, zero net magnetization occurs at a compensation temperature, denoted as $T^*$, which can be smoothly modulated by chemical composition. While amorphous materials have dominated the study of rare-earth transition-metal compensated ferrimagnets, crystalline compounds have received comparatively less attention. Unlike amorphous materials, crystalline materials host well-defined Bloch energy bands, characterized by Berry curvatures capable of generating large intrinsic anomalous Hall effects. Consequently, the search and design of materials with large Berry curvature has also emerged as a key focal point in spintronics research~\cite{fujitaGaugeFieldsSpintronics2011}.

Here we report the realization of compensated ferrimagnetism in a rare-earth transition-metal intermetallic compound, Tb(Mn$_{1-x}$Cr$_x$)$_6$Sn$_6$. Tb(Mn$_{1-x}$Cr$_x$)$_6$Sn$_6$ belongs to the $RT_6X_6$ ($R$ = rare earth, $T$ = transition metal, $X$ = Si, Ge, Sn) family, known for its kagome lattice structure and intriguing band structure features such as flat bands, van Hove singularities, and Dirac points~\cite{YMS_arpes, GVSHVS_arpes, GVS_Kdoping}. Notably, the ferrimagnet \ch{TbMn6Sn6} exhibits a large intrinsic AHE, which was initially attributed to Berry curvature arising from a gapped 2D Dirac point above but near the Fermi level~\cite{quantumlimitmagnetism_TMS}. This picture, if correct, would preclude Cr doping (hole doping) as a knob to tune the net magnetization without decreasing the intrinsic AHE, since that would shift the Fermi level away from the Dirac point. Furthermore, the 2D nature of the Dirac point would limit the large AHE to one particular geometry. Fortunately, it was later shown~\cite{Binghai,Lee2023PRB,originofSR_TMS_PRB} that the AHE is accumulated over the entire Brillouin zone and does not depend dramatically on the Fermi level; it was also shown that the AHE in similar compounds is, in fact, 3D~\cite{3D}, which is also favorable for applications.

In this work, we show that the substitution of Cr for Mn reduces the magnetic moment of the transition metal site, which eventually turns Tb(Mn$_{1-x}$Cr$_x$)$_6$Sn$_6$ into a compensated ferrimagnet. Near compensation, we observed a divergence of the coercive field at low temperatures. Furthermore, contrary to the initially advocated 2D Dirac point scenario, Cr substitution significantly enhances intrinsic AHE, likely due to the tuning of the Fermi level to various Berry curvature hot spots~\cite{originofSR_TMS_PRB}. Our findings pave the way for the rational design of a compensated ferrimagnet with a large intrinsic AHE.

\begin{figure*}
    \centering
    \includegraphics[width=0.9\textwidth]{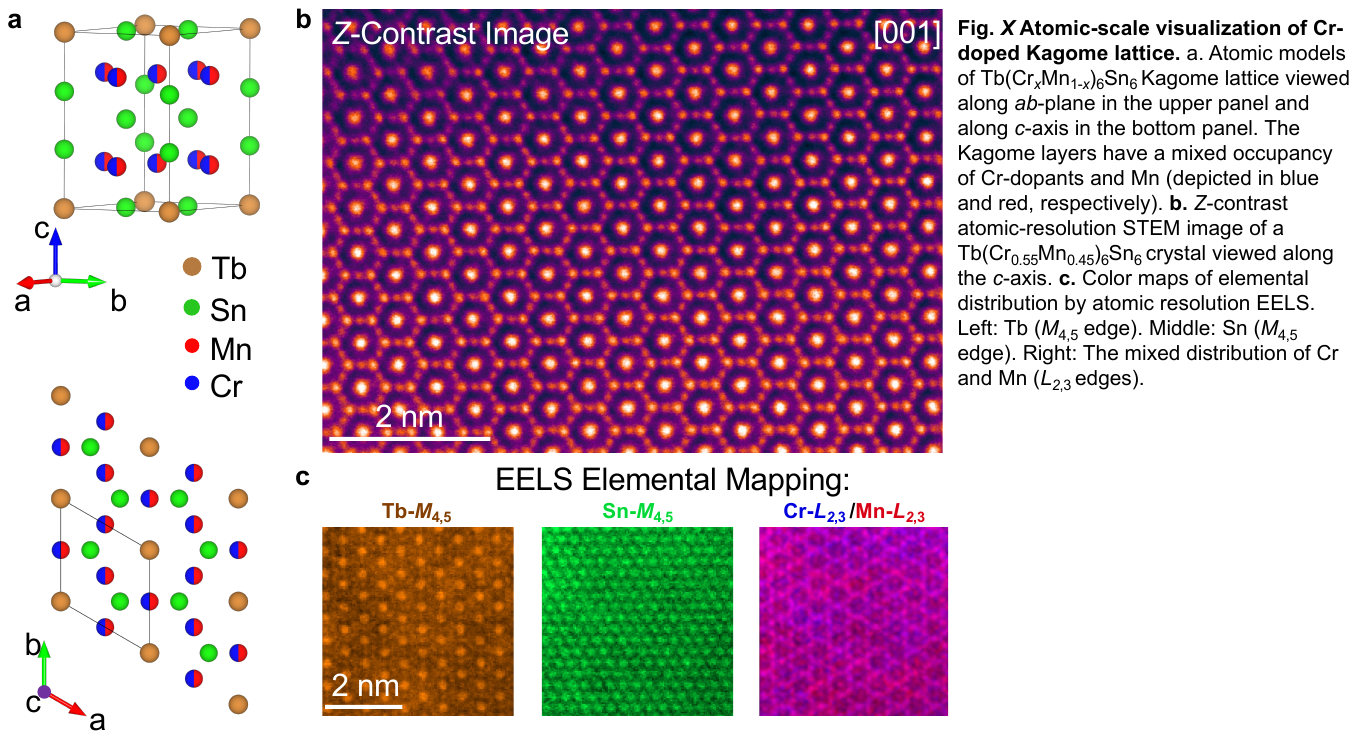}
    \caption{\textbf{Atomic-scale visualization of Tb(Mn$_{1-x}$Cr$_x$)$_6$Sn$_6$ kagome lattice.} \textbf{a,} Atomic models of the Tb(Mn$_{1-x}$Cr$_x$)$_6$Sn$_6$ lattice viewed along the \textit{ab}-plane in the upper panel and along the \textit{c}-axis in the bottom panel. The kagome layers have a mixed occupancy of Cr-dopants and Mn (depicted in blue and red, respectively). \textbf{b,} \textit{Z}-contrast atomic-resolution STEM image of an $x = 0.55$ crystal viewed along the \textit{c}-axis. \textbf{c,} Color maps of elemental distribution by atomic-resolution EELS. Left: Tb ($M_{4,5}$ edge), middle: Sn ($M_{4,5}$ edge), right: mixed distribution of Cr and Mn ($L_{2,3}$ edges).}
    \label{fig:STEMfigure}
\end{figure*}

\section{Basic characterization}

\begin{figure*}
    \centering
    \includegraphics[width=0.9\textwidth]{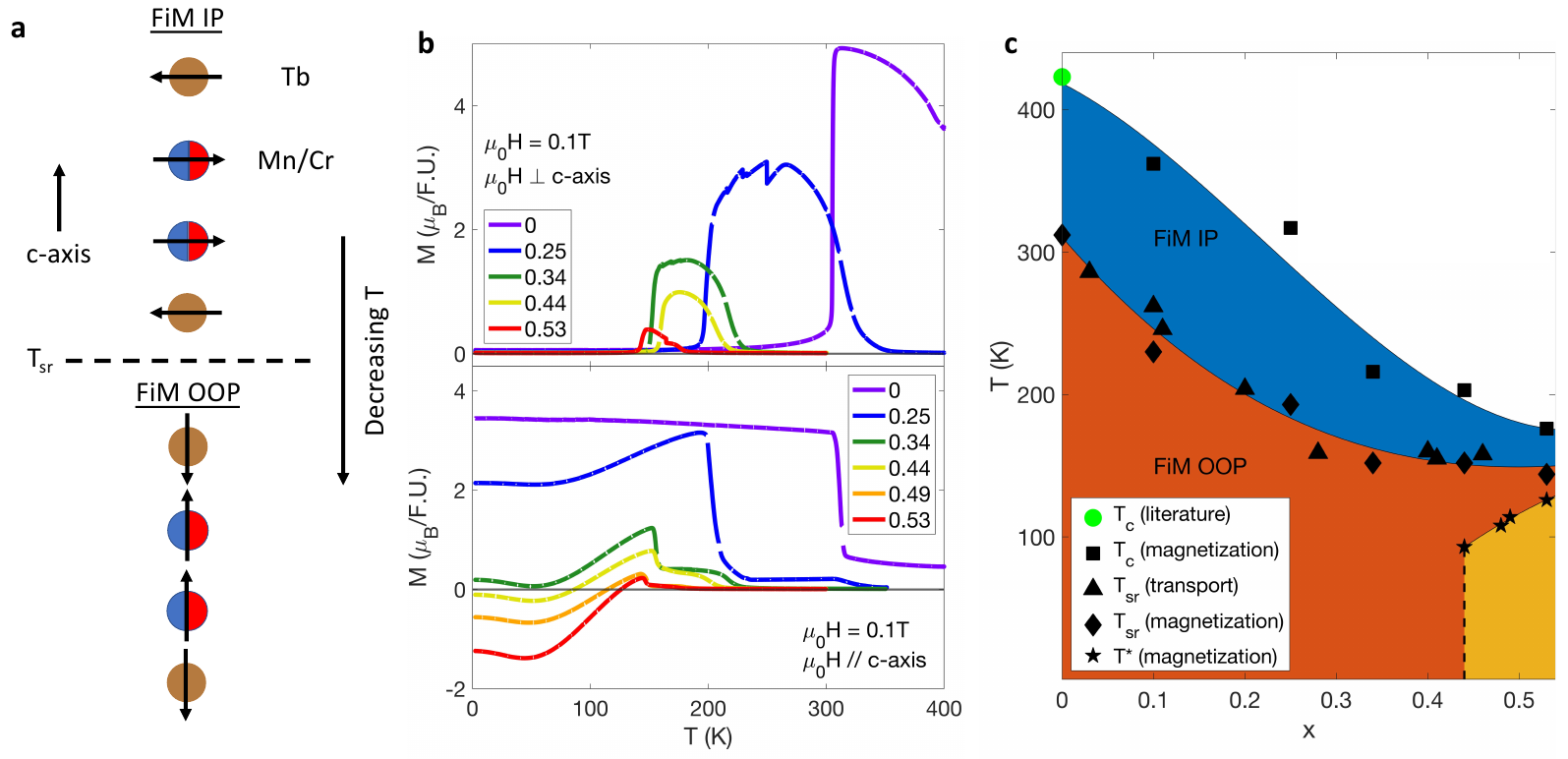}
    \caption{\textbf{Spin reorientation transition and the phase diagram of Tb(Mn$_{1-x}$Cr$_x$)$_6$Sn$_6$.} \textbf{a,} Orientation of magnetic moments of Tb(Mn$_{1-x}$Cr$_x$)$_6$Sn$_6$ in the high-temperature ferrimagnetic in-plane (FiM IP) (top) and low-temperature ferrimagnetic out-of-plane (FiM OOP) (bottom) phases. \textbf{b,} Low-field magnetization as a function of temperature of Tb(Mn$_{1-x}$Cr$_x$)$_6$Sn$_6$ for a variety of $x$ with an in-plane (top) or out-of-plane (bottom) magnetic field of \SI{0.1}{T} applied. The sudden decrease/increase of in-plane/out-of-plane magnetization indicates the spin reoirentation transition at $T_{sr}$. \textbf{c,} $x-T$ phase diagram of Tb(Mn$_{1-x}$Cr$_x$)$_6$Sn$_6$ as determined from magnetization and transport measurements.}
    \label{fig:MvTsandphasediagram}
\end{figure*}

High-quality single crystals of Tb(Mn$_{1-x}$Cr$_x$)$_6$Sn$_6$ with $x$ up to 0.55 were grown using a Sn flux method. The magnetization and magneto-transport properties of these crystals were measured by the vibration sample magnetometry (VSM) option of a Quantum Design Physical Property Measurement System (PPMS) Dynacool and standard AC electrical transport techniques, respectively. Further details can be found in the methods section. Fig.~\ref{fig:STEMfigure}a shows the crystal structure of \ch{TbMn6Sn6}, which consists of two layers of kagome lattices formed by Mn atoms and one layer of triangular lattice composed of Tb atoms, viewed along the ab-plane (top) and c-axis (bottom). This crystal structure is preserved with Cr doping, as exemplified in the \textit{Z}-contrast atomic-resolution scanning transmission electron microscope (STEM) image shown in Fig.~\ref{fig:STEMfigure}b where the ab-plane kagome motif can be observed in an $x = 0.55$ sample. Additionally, atomic-resolution electron energy-loss spectroscopy (EELS) was used to map out the elemental distribution in this sample which is presented in Fig.~\ref{fig:STEMfigure}c. Notably,  the Cr substitution for Mn is relatively homogeneous even at the nanometer scale.

\ch{TbMn6Sn6} orders ferrimagnetically at \SI{423}{K} ($T_c$)~\cite{TMS_phasetransitions2}. Within the layers the Mn moments and the Tb moments order ferromagnetically, but the Tb and Mn layers couple antiferromagnetically, resulting in a ferrimagnetic structure where Tb and Mn moments are antiparallel (schematically shown in the top portion of Fig.~\ref{fig:MvTsandphasediagram}a (FiM IP)). Neutron scattering studies have revealed that at base temperature, the size of Mn moment is 2.4 Bohr magnetons ($\mu_B$) and the size of Tb moment is 8.6$\mu_B$, leading to a net 5.8$\mu_B$ moment per formula unit along the direction Mn moments~\cite{TMS_phasetransitions2}.  Fig.~\ref{fig:MvTsandphasediagram}b shows the temperature dependence of the magnetization of Tb(Cr$_x$Mn$_{1-x}$)$_6$Sn$_6$ with a small in-plane (top) and out-of-plane (bottom) magnetic field of \SI{0.1}{T}. The ferrimagnetic ordering temperature of the $x = 0$ sample exceeds the instrument limit (\SI{400}{K}), thus it is not visible within this dataset. Nevertheless, a sudden decrease in in-plane magnetization and an increase in out-of-plane magnetization are clearly observed at \SI{309}{K} ($T_{sr}$). The abrupt change in magnetization is associated with a spin reorientation transition~\cite{quantumlimitmagnetism_TMS}, where the Tb and Mn moments switch from a high temperature in-plane orientation to a low temperature out-of-plane orientation (FiM OOP), as shown in the bottom portion of Fig.~\ref{fig:MvTsandphasediagram}a. This transition can be attributed to the competition between the c-axis uniaxial anisotropy favored the Tb moments and the easy-plane anisotropy preferred by the Mn moments. At high temperatures, the Tb moments fluctuate more due to a weaker exchange interaction, allowing the easy-plane anistotropy of the Mn moments dominates. At low temperatures, the scenario is reversed, and the c-axis uniaxial anisotropy of the Tb moments prevails~\cite{originofSR_TMS_PRB, orbitalcharacter_TMS}.

As shown in Fig.~\ref{fig:MvTsandphasediagram}b, the temperature dependence of the in-plane magnetization for Cr doped samples closely resemble that of the parent compound except that $T_c$, $T_{sr}$ and the magnitude of the magnetization steadily decrease with increasing $x$. This is consistent with a previous study on polycrystalline samples, where Cr doping reduced the average moment size of the transition metal, thereby suppressing the overall ferrimagnetic ordering temperature~\cite{polycrystallineTCMS_Buschow}. However, the out-of-plane magnetization exhibits a more intriguing behavior. While the features at $T_c$ and $T_{sr}$ are consistent with the in-plane magnetization data, the out-of-plane magnetization demonstrates a strong temperature dependence below $T_{sr}$, contrasting with the nearly constant magnetization observed in the parent \ch{TbMn6Sn6} over this temperature range. For the Cr doped samples, the out-of-plane magnetization decreases with temperature, reaching a local minimum near \SI{60}{K}. For $x$ greater than 0.4, the magnetization continues to decrease even after reaching zero at $T^*$, resulting in negative value for temperatures below $T^*$. This anti-alignment of the moment with the external magnetic field has been observed in other ferrimagnets driven to magnetic compensation by changing temperature~\cite{GdCrO3_Tstar}.

Using the magnetization data presented here and the resistivity data presented in Supplementary Note 1, along with the literature value of $T_{c}$ for $x = 0$ ~\cite{TMS_phasetransitions1, TMS_phasetransitions2}, an $x-T$ phase diagram was constructed and is shown in Fig.~\ref{fig:MvTsandphasediagram}c. It can be seen that both $T_c$ and $T_{sr}$ decrease rapidly with increasing $x$ in the low $x$ regime, but become relatively constant with further doping above approximately $x = 0.3$. For higher Cr concentration, $T^*$ emerges, delineating the boundary of a region where the magnetization is anti-aligned with the external field during the cool-down process. In the following sections, we will examine the hysteresis loop of both the magnetization and the anomalous Hall resistivity as a function of doping and temperature. Our results reveal that this unusual behavior arises from magnetic compensation driven by Cr doping and thermal fluctuations of Tb moments.

\begin{figure*}
    \centering
    \includegraphics[width=0.9\textwidth]{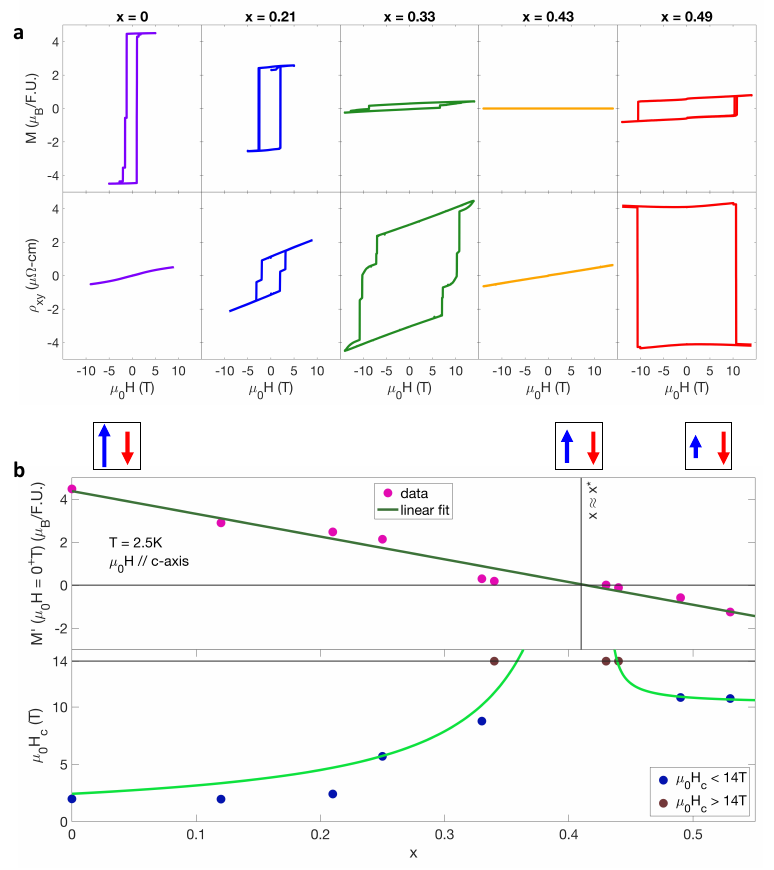}
    \caption{\textbf{Magnetic compensation and giant coercivity through chemical doping.} \textbf{a,} Magnetization (top row) and $\rho_{xy}$ (bottom row) as a function of magnetic field for a variety of $x$ at \SI{2.5}{K}. Plots in the same column are from the same $x$ which is indicated by the text on top of the column. \textbf{b,} Top: M' (described in the main text) as a function of $x$ at \SI{2.5}{K} with decreasing magnetic field. Bottom: $\mu_0H_c$ as a function of $x$ from magnetization measurements.}
    \label{fig:compensationthroughdoping}
\end{figure*}

\section{Chemical doping induced magnetic compensation and giant coercivity}

 We begin by examining the doping dependence at the base temperature. Fig.~\ref{fig:compensationthroughdoping}a shows the magnetization (top row) and $\rho_{xy}$ (bottom row) as functions of the magnetic field parallel to the c-axis at \SI{2.5}{K} for five representative Cr concentrations. The magnetization of Tb(Mn$_{1-x}$Cr$_x)_6$Sn$_6$ shows a clear hysteresis loop characteristic of a hard ferrimagnet with uniaxial anisotropy. As the Cr concentration increases, the saturated magnetization decreases, reaching a value near zero at a composition $x^* \approx 0.43$
 , before increasing again for $x > x^*$. The reduction in saturated magnetization is accompanied by a significant increase in the coercive field. A previous neutron scattering study on low-$x$ Tb(Mn$_{1-x}$Cr$_x$)$_6$Sn$_6$ revealed that Cr doping at the Mn site reduces the moment on that site without substantially changing the moment on the Tb site~\cite{polycrystallineTCMS_Buschow}. Extending this observation to higher dopings suggests that at $x^*$ the total magnetization from the Mn/Cr and Tb moments becomes equal, resulting in a net magnetization of zero. As $x$ increases beyond $x^*$, the total magnetization of the Tb moments surpasses that of the Mn/Cr moments, causing the net magnetization to rise again, aligning with the Tb moment. 

To confirm this hypothesis, we plot $M'$ in the top panel of Fig.~\ref{fig:compensationthroughdoping}b, where $M' = M$ for $x < x^*$ and $M' = -M$ for $x > x^*$. Here $M$ represents the experimentally measured zero field magnetization after the application of a large positive field to polarized the moments. The assumption is that $M'$ corresponds to the net magnetization of the Mn/Cr moments minus the net magnetization of the Tb moments. Indeed, $M'$ decreases roughly linearly as a function of $x$ while (by construction) crossing zero at $x^*$. A linear fit of these data indicates that the average moment decreases by roughly 1.8$\mu_B$ per Cr substitution, in good agreement with the previous neutron scattering study~\cite{polycrystallineTCMS_Buschow}. Since each Mn ion has a moment of 2.4$\mu_B$ in \ch{TbMn6Sn6}, this implies that each Cr ion in Tb(Mn$_{1-x}$Cr$_x$)$_6$Sn$_6$ carries a moment of 0.6$\mu_B$ if the Mn moments remain the same regardless of doping. This is quite similar to the value of 0.48$\mu_B$ which was found to be the moment per Cr ion in \ch{TbCr6Ge6} via neutron scattering measurements~\cite{TCG_neutron}. Thus, Cr is likely in a low-spin state in these materials. 

To gain further insight, we performed density functional theory (DFT) calculations to investigate the magnetic state of Tb(Mn$_{1-x}$Cr$_x$)$_6$Sn$_6$. Interestingly, we found that for TbCr$_2$Mn$_4$Sn$_6$ (with both DFT+U frozen core Tb$^{3+}$) the lowest energy states correspond to Cr forming one-dimensional in-plane antiferromagnetic chains. Thus, a potential major contribution to the magnetic moment reduction with Cr doping is due neither to dilution of the Mn sublattice by nonmagnetic ions, nor to suppression of Mn moments, but by the tendency of Cr appearing next to each other in the plane to form antiferromagnetic clusters. The results of these calculations are shown in Supplementary Note 2.

The bottom panel of Fig.~\ref{fig:compensationthroughdoping}b shows the coercive field, $\mu_0H_c$, extracted from the magnetization measurement. The coercive field approximately follows the inverse of the zero-field magnetization, consistent with the flipping of the magnetization when the Zeeman energy $\mu_0H_cM$ overcomes the domain depinning energy~\cite{compensatedheusler, FeTb_giantcoercivity}. Near $x^*$ the coercive field exceeds \SI{14}{T}, the maximum field achievable by the PPMS used. To our knowledge, this is the largest coercive field ever observed in a crystalline material~\cite{coercivefieldreview}. This massive coercivity is a result of the small magnetic moment observed combined with the large magnetocrystalline anisotropy of the Tb ions at low temperatures~\cite{originofSR_TMS_PRB, TVS_rosenberg}. These observations are consistent with Tb(Mn$_{1-x}$Cr$_x$)$_6$Sn$_6$ behaving as a nearly compensated ferrimagnet, exhibiting a giant coercive field at low temperatures between $0.30 \lesssim x \lesssim 0.45$, where the net Tb moments and the net Mn/Cr moments nearly cancel each other. 


In contrast to magnetization, $\rho_{AH}$ increases with Cr concentration. For samples very close to $x^*$, the hysteresis in $\rho_{AH}$ cannot be detected by magnetic field sweep at base temperature simply because the coercive field exceeds \SI{14}{T}. However, when the sample is field-cooled from high temperature, a large $\rho_{AH}$ can still be observed even as $M$ approaches zero. For $x > x^*$, the anomalous Hall effect switches sign, i.e. $\rho_{AH}$ switches from positive to negative when the magnetization is positive. Given that the energy bands near the Fermi level are primarily composed of transition metal d-orbitals, the sign change of is consistent with the anomalous Hall effect being driven by the transition metal sites. In-depth analysis of the anomalous Hall effect will be presented in Sec.~\ref{sec:AHE}.

\section{Temperature induced magnetic compensation}

\begin{figure*}
    \centering
    \includegraphics[width=0.9\textwidth]{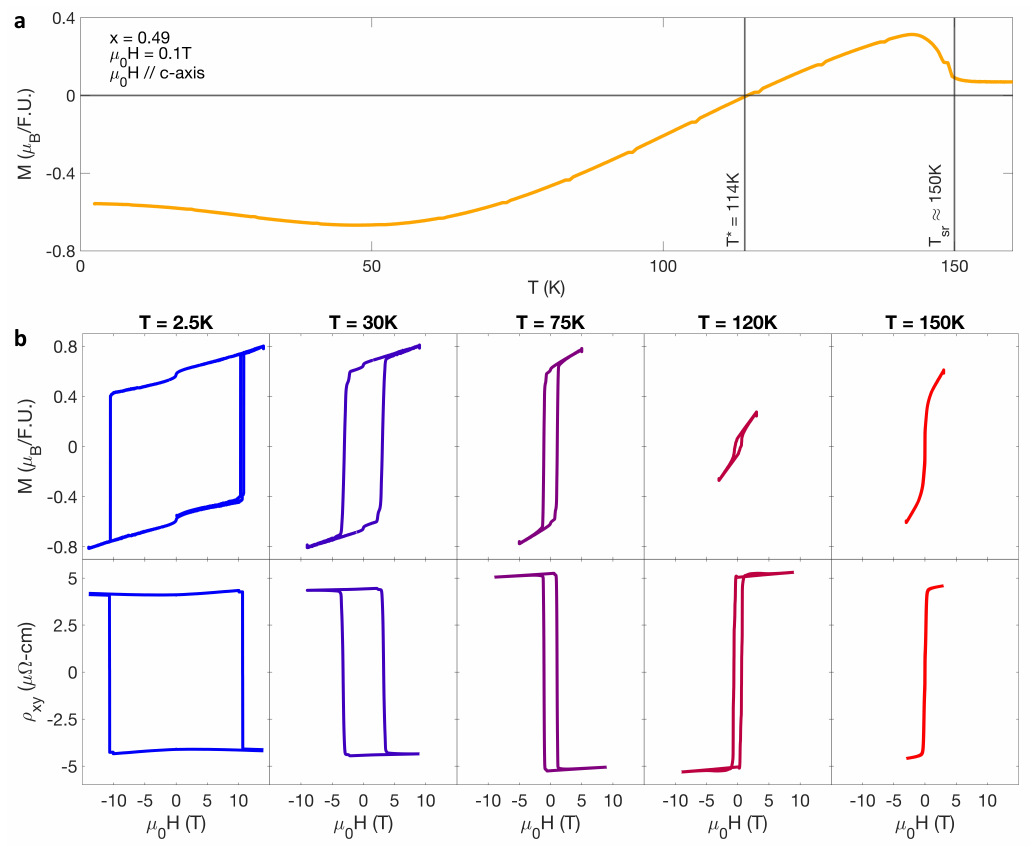}
    \caption{\textbf{Temperature dependence of magnetization and sign reversal of $\rho_{AH}$.} \textbf{a,} Low-field magnetization with an out-of-plane magnetic field for $x = 0.49$ as a function of temperature. \textbf{b,} Magnetization (top row) and $\rho_{xy}$ (bottom row) as a function of magnetic field at a variety of temperatures for the same sample. Plots in the same column are taken at the same temperature which is indicated by the text on top of the column.}
    \label{fig:p49tempdep}
\end{figure*}

In addition to chemical substitution, temperature is another effective parameter for controlling magnetic compensation. Fig.~\ref{fig:p49tempdep}a displays the previously shown out-of-plane magnetization as a function of temperature for a sample with $x =0.49$. This data was collected as the sample is cooled under a \SI{0.1}{T} c-axis magnetic field. At the spin reorientation transition temperature $T_{sr} =$ \SI{150}{K}, the magnetization abruptly increases as all moments reorient from in-plane to out-of-plane. Below $T_{sr}$, the magnetization reaches a maximum and then decreases with decreasing temperature. It crosses zero at $T^* =$ \SI{114}{K} and remains negative for temperatures below $T^*$. 

To gain a deeper understanding of this behavior, we measured the field dependence of magnetization and Hall resistivity $\rho_{xy}$ across a range of temperatures, as shown in the top and bottom rows of Fig.~\ref{fig:p49tempdep}b, respectively. Each column represents data taken at the same temperature, which is noted above the column. In the \SI{2.5}{K} data, in addition to the large hysteresis loop, a small additional paramagnetic contribution can be observed near $\mu_0H =$ \SI{0}{T} in the magnetization versus field data. This contribution is highly sample dependent and only noticeable in samples with $x$ greater than 0.4, suggesting it may result from either disordered spins or a minor impurity phase. As the temperature increases, $\mu_0H_c$ decreases significantly (the temperature dependence of $\mu_0H_c$ for multiple samples is detailed in Supplementary Note 3). Near $T^*$ ($\approx$ \SI{120}{K}), the hysteresis loop in $\rho_{xy}$ reverses direction, whereas the magnetization vanishes. As temperature increases above $T^*$, the magnetization increases again. The temperature dependence of the magnetization and the anomalous Hall effect mirrors the doping dependence at base temperature, indicating that the magnetic compensation can also be induced by temperature sweep. Instead of a reduction in average Mn/Cr moments due to Cr substitution, the magnetic compensation at $T^*$ is driven by the reduction of the average Tb moments at elevated temperatures due to thermal fluctuations, which is the same mechanism that drives the spin-reorientation transition~\cite{originofSR_TMS_PRB}. Such temperature-driven compensation is commonly seen in other rare-earth transition metal alloy system, where sign switching of $\rho_{AH}$ has been observed~\cite{reversalofAHE}.

With these new insights, we can now explain the magnetization reversal observed in the field-cooled measurements. In samples with $x > \approx 0.4$, the Mn/Cr moments have decreased sufficiently that the Tb moments can match them in magnitude at $T^*$. As the temperature continues to decrease in these samples, the net Tb moment becomes larger in magnitude than the Mn/Cr moment, causing the net magnetization to become anti-aligned with the external field. Additionally, the coercive field rapidly increases with decreasing temperature and exceeds the external field below $T^*$. Consequently, the magnetization becomes trapped in the reverse configuration, leading to an overall negative value of magnetization.


\section{Enhancement of anomalous Hall effect with doping}
\label{sec:AHE}


\begin{figure*}
    \centering
    \includegraphics[width=0.9\textwidth]{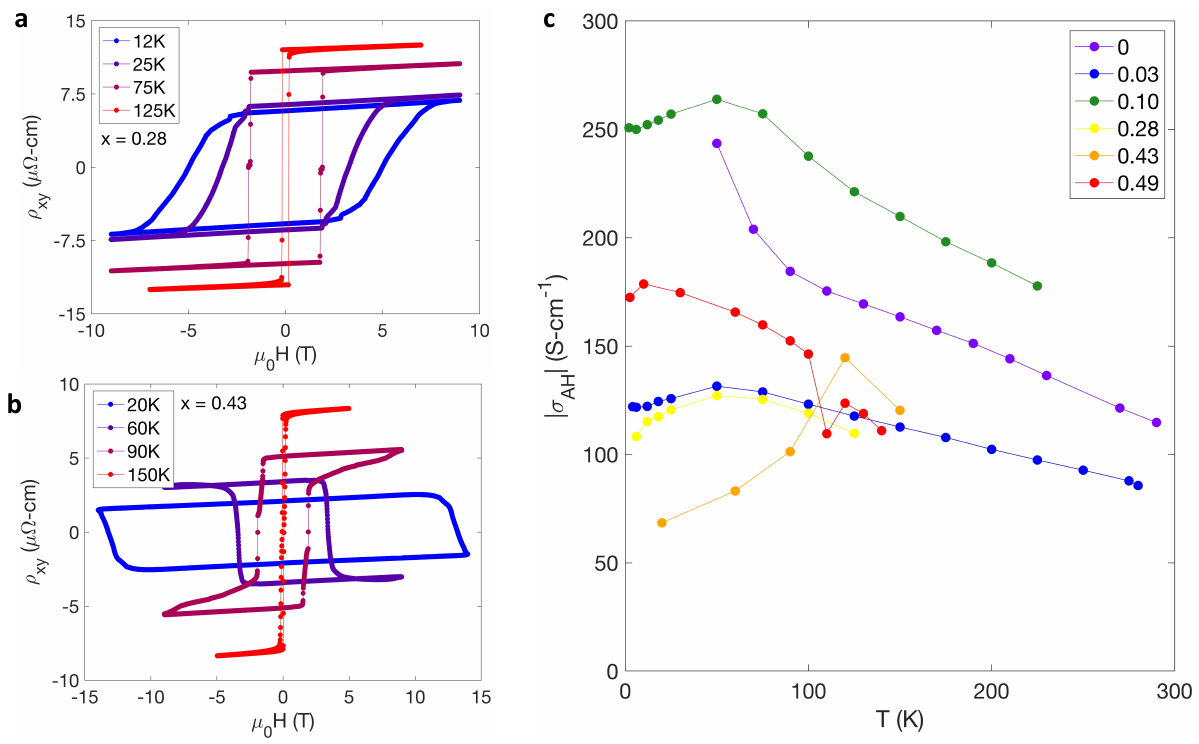}
    \caption{\textbf{Anomalous Hall effect of Tb(Mn$_{1-x}$Cr$_x$)$_6$Sn$_6$.} \textbf{a, b,} $\rho_{xy}$ as a function of $\mu_0H$ at several temperatures for $x = $ 0.28 and 0.43, respectively. \textbf{c,} Anomalous Hall conductivity $|\sigma_{AH}|$ as a function of temperature for a variety of dopings of Tb(Mn$_{1-x}$Cr$_x$)$_6$Sn$_6$.}
    \label{fig:basicHall}
\end{figure*}

\begin{figure*}
    \centering
    \includegraphics[width=0.9\textwidth]{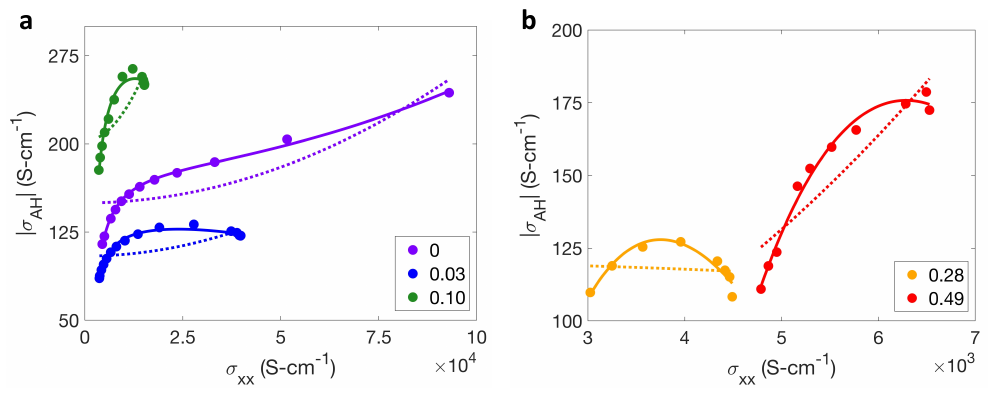}
    \caption{\textbf{Fitting of anomalous Hall conductivity.} \textbf{a, b,} $|\sigma_{AH}|$ as a function of $\sigma_{xx}$ for low (a) and high (b) $x$. Dashed lines correspond to fits utilizing the standard anomalous Hall fitting function and solid lines show fits which include the extra term described in the main text.}
    \label{fig:AHE}
\end{figure*}

\begin{figure*}
    \centering
    \includegraphics[width=0.9\textwidth]{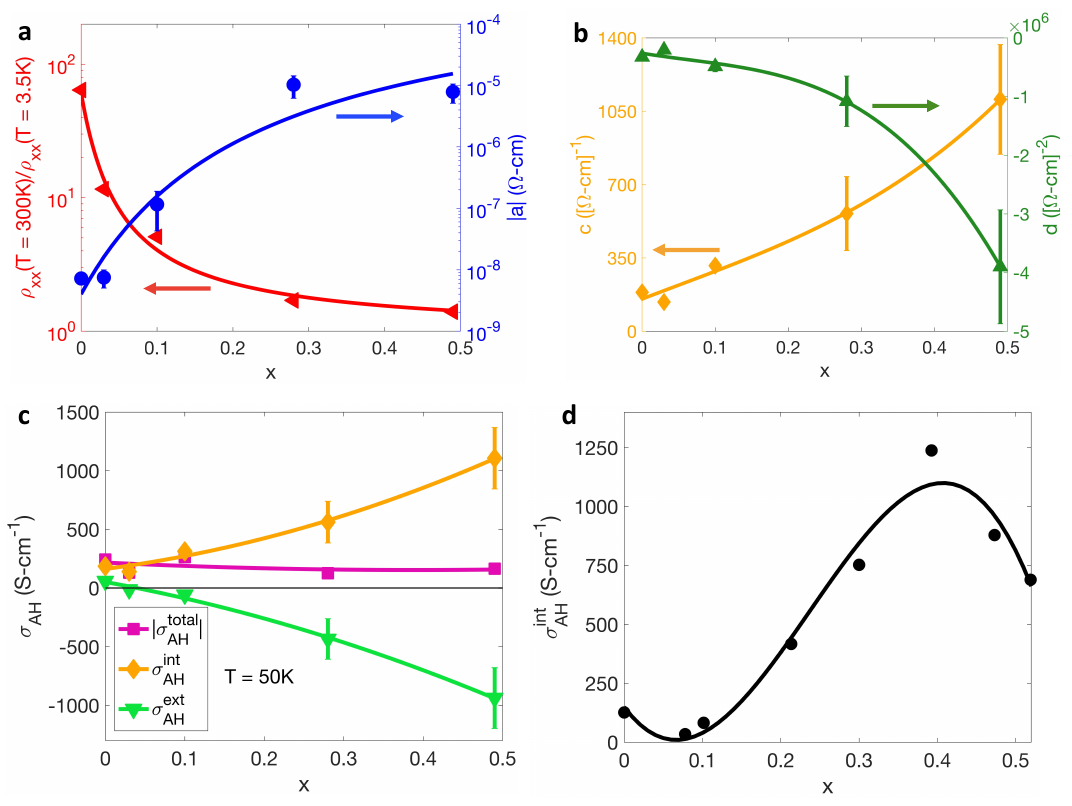}
    \caption{\textbf{Doping dependence of intrinsic and extrinsic anomalous Hall effect parameters.} \textbf{a,} Residual resistivity ratio and fit parameter $a$ (extrinsic contribution) as a function of $x$. \textbf{b,} Fit parameters $c$ (intrinsic contribution) and $d$ (spin fluctuations contribution) as a function of $x$. \textbf{c,} $|\sigma_{AH}|$ and extracted $\sigma_{AH}^{int}$ and $\sigma_{AH}^{ext}$ at \SI{50}{K} as a function of $x$. \textbf{d,} Calculated $\sigma_{AH}^{int}$ as a function of $x$ using a rigid band model.}
    \label{fig:AHEfitparameters}
\end{figure*}

\begin{table}[h]
    \caption{Extracted parameters from conductivity fitting of AHE.}
    \centering
\begin{tabular}{||c | c | c | c | c||} 
 \hline
 $x$ & a ($\Omega$-cm) & c ([$\Omega$-cm]$^{-1}$) & d ([$\Omega$-cm]$^{-2}$) \\ [0.5ex] 
 \hline\hline
 0 & (7.21$\pm$0.64)$\times$10$^{-9}$ & 186$\pm$2.8 & (-3.26$\pm$0.22)$\times$10$^{5}$ \\ 
 \hline
 0.03 & (-7.48$\pm$2.45)$\times$10$^{-9}$ & 140$\pm$3.5 & (-2.04$\pm$0.18)$\times$10$^{5}$ \\
 \hline
 0.10 & (-1.16$\pm$0.73)$\times$10$^{-7}$ & 313$\pm$20.4 & (-4.87$\pm$0.88)$\times$10$^{5}$  \\
 \hline
 0.28 & (-1.03$\pm$0.41)$\times$10$^{-5}$ & 562$\pm$177 & (-1.08$\pm$0.43)$\times$10$^{6}$   \\ 
 \hline
 0.49 & (-7.86$\pm$2.72)$\times$10$^{-6}$ & 1107$\pm$262 & (-3.90$\pm$0.97)$\times$10$^{6}$  \\ [1ex] 
 \hline
\end{tabular}
    \label{tab:AHEcondparameters}
\end{table}
\begin{center}
\end{center}


We now turn to a more quantitative analysis of the anomalous Hall effect. Fig.~\ref{fig:basicHall}a and b show the field dependence of Hall resistivity $\rho_{xy}$ of samples with $x =$ 0.28 and 0.43 measured at various temperatures. Similar plots of samples with other Cr concentrations can be found in Supplemental Note 4. All samples exhibit pronounced hysteresis and a well-defined anomalous Hall effect at zero field. For higher Cr concentrations, the anomalous Hall effect changes sign as a function of temperature as a result of magnetic compensation, as discussed in the previous section.

We calculated the anomalous Hall conductivity, $\sigma_{AH}$, using the relation $\sigma_{AH} = {-\rho_{AH}}/({\rho_{xx,0}^2 + \rho_{AH}^2})$, where $\rho_{xx,0}$ is the zero-field longitudinal resistivity. Fig.~\ref{fig:basicHall}c presents the anomalous Hall conductivity as a function of temperature for various Cr concentrations. To facilitate comparison across different samples without the influence of sign switching, we plot only the absolute value $|\sigma_{AH}|$. The results for $x = 0$ are consistent with previous reports \cite{originofSR_TMS_PRB}. Cr doping does not significantly alter $|\sigma_{AH}|$, which changes by less than an order of magnitude across all measured Cr concentrations. Additionally, this quantity is not strongly dependent on temperature.

There are two contributions to the anomalous Hall effect: the intrinsic and extrinsic contributions. The former is originated from the Berry curvature of the band structure, while the latter is due to the skew scattering and side-jump mechanisms. Separating the intrinsic and extrinsic contributions of the anomalous Hall effect is a challenging task. Traditionally, this is done by fitting the anomalous Hall conductivity versus longitudinal conductivity to the function $\sigma_{AH}=a\sigma_{xx}^2+c$, where $c = \sigma_{AH}^{int}$ is the intrinsic contribution, and the $a\sigma_{xx}^2$ term corresponds to extrinsic skew scattering and side jump contributions~\cite{properscalingAHE}. However, it has been shown that this function does not adequately fit the anomalous Hall conductivity in TbMn$_6$Sn$_6$. An additional non-standard term, 
tentatively associated with spin fluctuations as discussed in Ref.~\cite{originofSR_TMS_PRB}, can be included to achieve a better fit, leading to an equation of the following form:

\begin{equation}
    \sigma_{AH}=a\sigma_{xx}^2+c+{d}\sigma_{xx}^{-1}
    \label{eqn:AHEcon}
\end{equation}

We fitted the $\sigma_{AH}$ versus $\sigma_{xx}$ data for each sample, as shown in Fig.~\ref{fig:AHE}a and b for low and high doping levels, respectively, to extract $\sigma_{AH}^{int}$ and $\sigma_{AH}^{ext}$. The dashed lines represent fits to the standard anomalous Hall equation, while the solid lines include an additional term proportional to $\frac{1}{\sigma_{xx}}$ (Eqn.~\ref{eqn:AHEcon}). This additional term decisively improves the fit, especially for higher dopings. 

The doping dependence of the fit parameters are presented in Table~\ref{tab:AHEcondparameters} and plotted in Fig.~\ref{fig:AHEfitparameters}. It can be seen that the $a$ coefficient associated with the skew scattering changes sign and increases dramatically as Cr concentration increases. This can be explained by the increase of disorder scattering due to Cr doping, and it is consistent with the substantial drop of residual resistance ratio (RRR), as shown in Fig.~\ref{fig:AHEfitparameters}a. The extracted non-standard parameter ($d$) increases roughly an order of magnitude over the measured dopings as shown in Fig.~\ref{fig:AHEfitparameters}b, which is consistent with the enhanced Tb fluctuations corresponding to reduction of $T_c$ with increasing $x$. Also shown in this panel is a substantial enhancement of the extracted parameter $c$ ($=\sigma_{AH}^{int}$) in orange. To further validate these fits, analyses using the resistivity form of the AHE equations are detailed in Supplementary Note 4 and yield similar results.

This extracted $\sigma_{AH}^{int}$ with increasing $x$ is plotted again in orange in Fig.~\ref{fig:AHEfitparameters}c. Also shown is a simultaneous development of a large negative $\sigma_{AH}^{ext}$ with increasing $x$ (results shown for data taken at \SI{50}{K}). Due to the opposite signs of the intrinsic and extrinsic contributions, the overall $|\sigma_{AH}^{total}|$ does not vary significantly with doping, and thus can be completely missed barring a proper scaling analysis.

\section{Discussion}

\ch{TbMn6Sn6} has attracted significant attention due to reports of a large intrinsic anomalous Hall effect, initially attributed to quasi two-dimensional gapped Dirac points, a characteristic feature of the kagome lattice band structure~\cite{quantumlimitmagnetism_TMS}. However, this interpretation has been challenged by recent calculations, which suggest that the relevant Dirac points are far from the Fermi level and that the intrinsic anomalous Hall effect primarily arises from the anti-crossing of energy bands at other locations in momentum space~\cite{originofSR_TMS_PRB,compensatedheusler,Lee2023PRB}. These studies further indicate that intrinsic anomalous Hall conductivity could be significantly enhanced if the material is hole-doped, a condition that can be achieved through Cr substitution.

The pronounced doping dependence of $\sigma_{AH}^{int}$ observed in this work sheds more light on this debate. According to the gapped Dirac point model proposed in Ref.~\cite{quantumlimitmagnetism_TMS}, $\sigma_{AH}^{int}$ should decrease with Cr doping, as Cr doping lowers the Fermi level away from the gapped Dirac points, where the Berry curvature is concentrated. In contrast, the observed increase in $\sigma_{AH}^{int}$ with Cr doping (i.e., hole doping) aligns better with scenarios proposed in Refs.~\cite{originofSR_TMS_PRB,compensatedheusler,Lee2023PRB}. To verify this, we calculate $\sigma_{AH}^{int}$ as a function of hole doping by rigid band shift the Fermi level of \ch{TbMn6Sn6}, as shown Fig.~\ref{fig:AHEfitparameters}d. The good agreement between the experiment (orange curve in Fig.~\ref{fig:AHEfitparameters}c) and the calculation (Fig.~\ref{fig:AHEfitparameters}d) unambiguously confirms that multiple anti-crossing features in the band structure contribute to the intrinsic anomalous Hall effect.

In summary, we successfully grew Cr-doped single crystals of \ch{TbMn6Sn6} and studied their magnetic and magnetotransport properties. Cr doping tunes the ferrimagnetic state towards magnetic compensation with a giant coercive field. Additionally, with increased Cr doping, $\sigma_{AH}^{int}$ becomes significantly larger, in strong agreement with first-principles calculations. This work offers a pathway for identifying and synthesizing compensated ferrimagnets with large intrinsic anomalous Hall conductivities.

\section*{Methods}

Single crystals of Tb(Mn$_{1-x}$Cr$_x$)$_6$Sn$_6$ were synthesized utilizing a flux method similar to those previously reported for \ch{TbMn6Sn6}~\cite{quantumlimitmagnetism_TMS, originofSR_TMS_PRB}. Mixtures of Tb (99.999\%) pieces, Cr powder (99.99\%), Mn powder (99.95\%), and Sn shot (99.999\%) were loaded into Canfield crucible sets~\cite{Canfield2016} with atomic ratios Tb:(Cr, Mn):Sn 2.5:15:97.5, then vacuum sealed in quartz tubes. These were heated up to 1150$^\circ$C in 12 hours, held at this temperature for 12 hours, then cooled to 600$^\circ$C in 150 hours. Then the growths were decanted in a centrifuge to separate the excess flux from the crystals. The doping concentration \textit{x} of each crystal of Tb(Mn$_{1-x}$Cr$_x$)$_6$Sn$_6$ used for measurements was determined using energy-dispersive X-ray spectroscopy (EDX) with a Sirion XL30 scanning electron microscope. Each crystal was polished prior to EDX measurements to remove residual flux on the surface of the crystal, and at least 8 points on each crystal was measured. The measured spread in \textit{x}$_{EDX}$ for each crystal used in this study (and most of the Tb(Mn$_{1-x}$Cr$_x$)$_6$Sn$_6$ in general) was limited to a few hundredths, indicating the doping is reasonably homogeneous throughout the crystal. A plot of \textit{x}$_{EDX}$ as a function of \textit{x}$_{nom}$ (the nominal $\frac{Cr}{Cr+Mn}$ included in a growth) for a number of growths is presented in Supplementary Note 5 along with other EDX information. Throughout this paper the measured \textit{x}$_{EDX}$ is referred to as \textit{x} for simplicity. Using the growth technique outlined here, crystals of Tb(Mn$_{1-x}$Cr$_x$)$_6$Sn$_6$ with 0 $<x<$ 0.55 were able to be grown. Some crystals with $x \gtrsim 0.3$ had significant Sn inclusions that could be detected via transport measurements (shorting of the resistivity below Sn's superconducting transition) and/or seen under a microscope. Measurements performed on these samples are not reported. It should be noted that for relatively high \textit{x}$_{nom}$ ($\geq$ 0.5) secondary phases of Cr (cubic crystals) and \ch{Tb3Sn7} (plate-like crystals) were also found in the growths, and growths with \textit{x}$_{nom}>$ 0.6 did not provide any crystals.

Transport measurements were performed on samples that were polished and cut by a wire saw to be bars with dimensions roughly \SI{1}{mm} $\times$ \SI{0.4}{mm} $\times$ \SI{0.05}{mm}. Silver paste and gold wires were used to make 5 point (Hall pattern) contacts. These measurements were performed in a Quantum Design Dynacool Physical Property Measurement System (PPMS) with standard lock-in techniques in temperatures ranging from \SI{1.7}{K} to \SI{400}{K} and in magnetic fields up to \SI{14}{T}. To eliminate any contributions from contact misalignment the in-line and Hall resistivities were symmetrized and anti-symmetrized with magnetic field, respectively.

Magnetization measurements were performed with the vibrating sample magnetometer option of the PPMS. Samples of the same size as the transport samples were attached to a quartz paddle with GE varnish such that the magnetic field was either out-of-plane (i.e. parallel to the c-axis) or in-plane.

Electron microscopy imaging and spectroscopy were carried out by an aberration-corrected scanning transmission electron microscopy (STEM) in a Nion UltraSTEM 100~\cite{Krivanek_2008}.  The electron microscope was operated with an acceleration voltage of 100 kV, using a semi-convergence angle of 32 mrad.  High-angle annular dark field (HAADF) images, also known as \textit{Z}-contrast images due to the intensity being proportional to atomic number,  were acquired with an inner (outer) semi-angle of 80 (200) mrad.  Electron energy-loss spectroscopy (EELS) data were acquired with a collection semi-angle of 48 mrad.

The DFT calculations were performed using a full-potential linear augmented plane wave (FP-LAPW) method, as implemented in \textsc{wien2k}~\cite{wien2k,wien2kU}.
Spin-orbit coupling (SOC) was included using a second variational method.
The generalized gradient approximation by Perdew, Burke, and Ernzerhof (PBE)~\cite{perdew1996} was used for the exchange-correlation potentials.
To generate the self-consistent potential and charge, we employed $R_\text{MT}\cdot K_\text{max}=8.0$ with Muffin-tin (MT) radii $R_\text{MT}=$ 2.8, 2.4, 2.4 and 2.5 atomic units (a.u.) for Tb, Mn, Cr and Sn, respectively.
The calculations are performed with 4800 $k$-points in the Full Brillouin zone (FBZ) and iterated until the charge differences between consecutive iterations are smaller than $10^{-4} e$ and the total energy differences are lower than $10^{-2}$ mRy/cell.
The strongly correlated Tb-$4f$ electrons are treated using the DFT+$U$ ($U$=10 eV) method with the fully-localized-limit (FLL) double-counting scheme~\cite{anisimov1993prb}.  
TbMn$_6$Sn$_6$ consists of six equivalent Mn atoms, which are replaced by one, two, or three Cr atoms depending on the doping concentration. We constructed one, three and three configurations by replacing on one, two and three Mn atoms to Cr atoms respectively and compared magnetic moments and total energies of each configuration.
We also performed virtual crystal (VC) calculations to confirm if the rigid band model is reliable in this study.
We kept lattice parameters the same as that of TbMn$_6$Sn$_6$ since the lattice parameters change is less than 0.5\% up to 33\% doping \cite{polycrystallineTCMS_Buschow}.

A realistic TB Hamiltonian was constructed using the maximally localized Wannier functions (MLWFs) method~\cite{marzari1997prb, souza2001prb, marzari2012rmp} implemented in \textsc{Wannier90}~\cite{mostofi2014cpc} after the self-consistent density-functional-theory calculations performed using \textsc{Wien2k}.  To circumvent complications associated with the 4$f$ state of the Tb atom while preserving the main physics, we employed an open-core method for the 4$f$ state.
A set of 118 Wannier functions (WFs) consisting of Tb-$5d$, Mn-$3d$, and Sn-$s,p$ orbitals offers an effective representation of the electronic structure near the Fermi level ($E_{\text{F}}$).
The intrinsic anomalous Hall conductivity (AHC) can be calculated by integrating the Berry curvature over the Brillouin zone (BZ)~\cite{Wang2006prb}:
\begin{eqnarray}
  \label{ahc:theo}
  \sigma_{\a\b}=-\dfrac{e^2}{\hbar} \int_\text{BZ}\dfrac{d\vec{k}}{(2\pi)^3}\sum_{n} f(E_{n\vec{k}}) \Omega_{n,\a\b}(\vec{k})\,,
\end{eqnarray}
where $f(E_{n\vec{k}})$ is the Fermi-Dirac distribution, $\Omega_{n,\a\b}(\vec{k})$ is the contribution to the Berry curvature from state $n$, and $\a,\b = \{x,y,z\}$.
A dense $256^3$ $k$-point mesh is used for the AHC calculations in TB.

\section*{Data availability}
All data supporting the findings of this study are available upon request.

\section*{Acknowledgments}

This work is supported by the Air Force Office of Scientific Research under grant FA9550-21-1-0068, the David and Lucile Packard Foundation and the Gordon and Betty Moore Foundation’s EPiQS Initiative, grant no. GBMF6759 to J.-H.C.. This material is based upon work supported by the National Science Foundation Graduate Research Fellowship Program under Grant No. DGE-2140004. Any opinions, findings, and conclusions or recommendations expressed in this material are those of the authors and do not necessarily reflect the views of the National Science Foundation. O.P. acknowledges support from the University of Washington Mary Gates Research Scholarship. This work was supported by NSF Grant. No. DMR-1719797. This work was supported by NSF Partnerships for Research and Education in Materials (PREM) Grant DMR-2121953. Part of this work was conducted at the Molecular Analysis Facility, a National Nanotechnology Coordinated Infrastructure site at the University of Washington which is supported in part by the National Science Foundation (grant NNCI-1542101), the University of Washington, the Molecular Engineering \& Sciences Institute, the Clean Energy Institute, and the National Institutes of Health. STEM imaging was conducted as part of a user project at the Center for Nanophase Materials Sciences (CNMS), which is a US DOE, Office of Science User Facility at Oak Ridge National Laboratory. Crystal structure images were generated with VESTA~\cite{VESTA}. This work was supported by the U.S.~Department of Energy, Office of Science, Office of Basic Energy Sciences, Materials Sciences and Engineering Division.
Ames Laboratory is operated for the U.S.~Department of Energy by Iowa State University under Contract No.~DE-AC02-07CH11358. I.I.M. acknowledges support from the National Science Foundation
under Award No. DMR-2403804.

\section*{Author contributions}
J.M.D., E.R., O.P, and K.H. grew the samples and performed the magnetization and magnetotransport measurements. G.R. performed the STEM experiments and analyzed the data with guidance from J.C.I.. Y.L., Z.N., L.K. and I.I.M. performed DFT calculations and contributed to the theoretical analysis. J.-H.C. oversaw the project. J.M.D., E.R., and J.-H.C. wrote the manuscript with input from all authors.

\section*{Competing interests}
The authors declare no competing interests.

\newpage

\bibliography{main}

\end{document}


\title{Supplementary Information for ``Giant coercivity and enhanced intrinsic anomalous Hall effect at vanishing magnetization in a compensated kagome ferrimagnet"}

\author{Jonathan M. DeStefano}
\affiliation{Department of Physics, University of Washington, Seattle WA, 98112, USA}
\author{Elliott Rosenberg}
\affiliation{Department of Physics, University of Washington, Seattle WA, 98112, USA}
\author{Guodong Ren}
\affiliation{Materials Sciences \& Engineering Department, University of Washington, Seattle WA, 98115, USA}
\author{Yongbin Lee}
\affiliation{Ames Laboratory, U.S. Department of Energy, Ames, Iowa 50011, USA}
\author{Zhenhua Ning}
\affiliation{Ames Laboratory, U.S. Department of Energy, Ames, Iowa 50011, USA}
\author{Olivia Peek}
\affiliation{Department of Physics, University of Washington, Seattle WA, 98112, USA}
\author{Kamal Harrison}
\affiliation{Department of Physics, University of Washington, Seattle WA, 98112, USA}
\affiliation{Department of Physics, University of Central Florida, Orlando FL, 32816, USA}
\affiliation{NanoScience Technology Center, University of Central Florida, Orlando FL, 32826, USA}
\author{Saiful I. Khondaker}
\affiliation{Department of Physics, University of Central Florida, Orlando FL, 32816, USA}
\affiliation{NanoScience Technology Center, University of Central Florida, Orlando FL, 32826, USA}
\affiliation{School of Electrical Engineering and Computer Science, University of Central Florida, Orlando, Florida 32826, USA}
\author{Liqin Ke}
\affiliation{Ames Laboratory, U.S. Department of Energy, Ames, Iowa 50011, USA}
\author{Igor I. Mazin}
\affiliation{Department of Physics and Astronomy and Quantum Science and Engineering Center, George Mason University, Fairfax, Virginia 22030, USA}
\author{Juan Carlos Idrobo}
\affiliation{Materials Sciences and Engineering Department, University of Washington, Seattle WA, 98115, USA}
\affiliation{Physical and Computational Sciences Directorate, Pacific Northwest National Laboratory, Richland WA, 99354, USA}
\author{Jiun-Haw Chu}
\email{jhchu@uw.edu}
\affiliation{Department of Physics, University of Washington, Seattle WA, 98112, USA}

\date{\today}

\maketitle


\section{Supplementary Note 1: Resistivity vs. temperature}

\begin{figure*}
    \centering
    \includegraphics[width=0.9\textwidth]{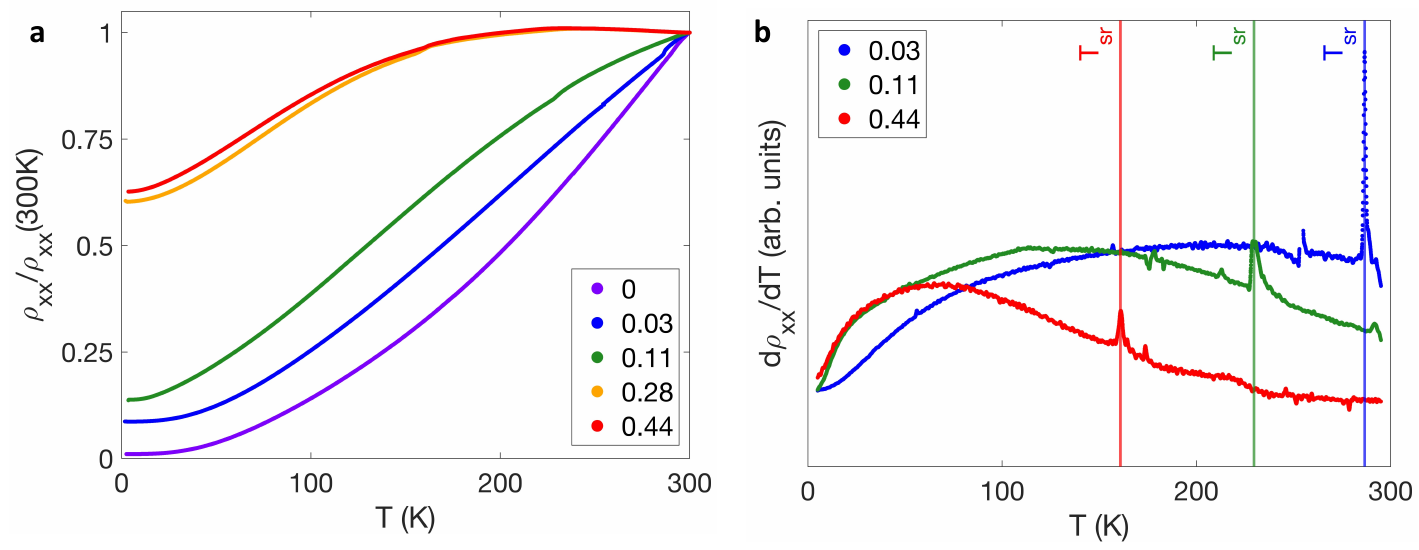}
    \caption{\textbf{Resistivity as a function of temperature of Tb(Mn$_{1-x}$Cr$_x$)$_6$Sn$_6$.} \textbf{a,} Normalized resistivity as a function of temperature for a several Tb(Mn$_{1-x}$Cr$_x$)$_6$Sn$_6$ samples. \textbf{b,} $\frac{d\rho_{xx}}{dT}$ as a function of temperature for Tb(Mn$_{1-x}$Cr$_x$)$_6$Sn$_6$ with $x = 0.44$. $T_{sr}$ is marked with a vertical line.}
    \label{fig:transport}
\end{figure*}

Fig.~\ref{fig:transport}(a) presents $\frac{\rho_{xx}}{\rho_{xx} (T = 300K)}$ as a function of temperature from 2-\SI{300}{K} for a variety of $x$ samples. Several samples only show data for temperatures down to \SI{3.6}{K} due to the emergence of small drops of resistivity below this temperature that arise from superconducting Sn that remains on the surface of the crystals. Any crystals with significant drops in the resistivity below this temperature were not used. The resistivity of $x = 0$ is consistent with previous reports~\cite{quantumlimitmagnetism_TMS, originofSR_TMS_PRB}, exhibiting metallic behavior below \SI{300}{K} and a residual resistivity ratio (RRR) of roughly 100. With increasing $x$ the RRR of Tb(Mn$_{1-x}$Cr$_x$)$_6$Sn$_6$ quickly decreases. For $x > \approx$0.25, non-metallic behavior (ie. $\frac{d\rho_{xx}}{dT} < $0) is observed at high temperatures. In these samples $\frac{d\rho_{xx}}{dT}$ switches sign at a temperature near $T_c$, but further work would be needed to fully understand this behavior. Small kinks can be seen in this data at $T_{sr}$ for $x > 0$ ($T_{sr}$ is not observed in these data for $x = 0$ since $T_{sr} >$ \SI{300}{K}) which is highlighted in Fig.~\ref{fig:transport}(b) which shows $\frac{d\rho_{xx}}{dT}$ as a function of temperature for several of the samples.

\section{Supplementary Note 2: DFT calculations of magnetization}

\begin{figure}[htb]
	\centering
	\begin{tabular}{c}
        \includegraphics[width=0.5\textwidth]{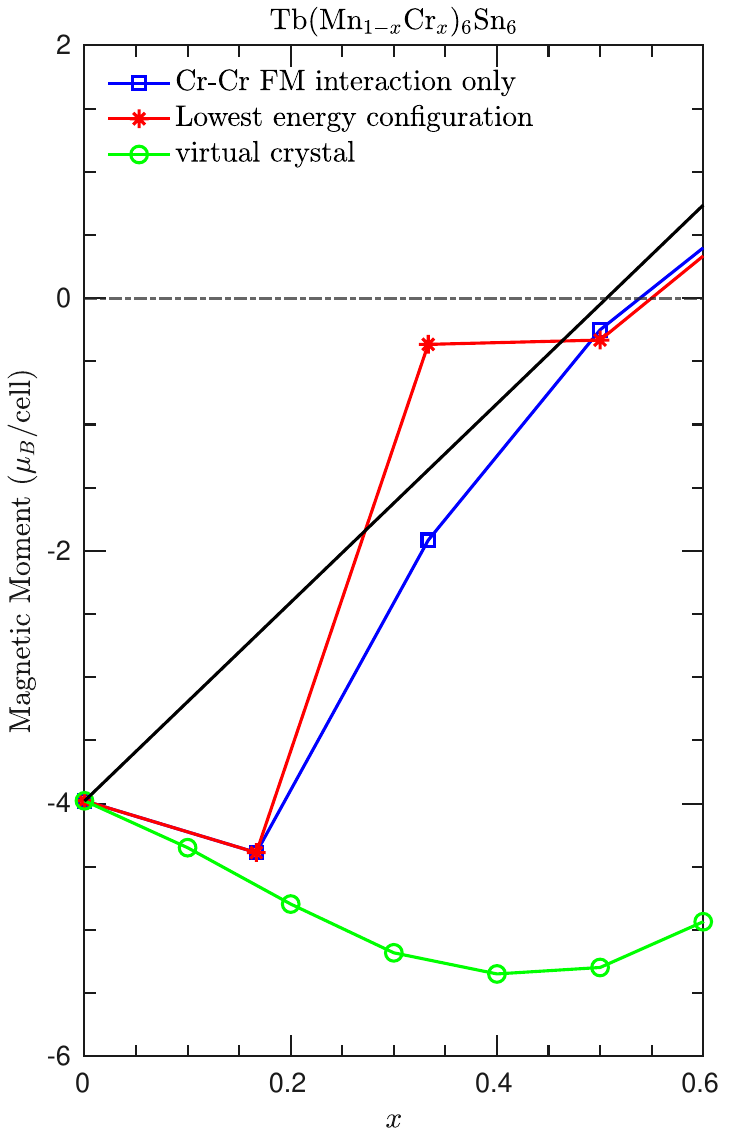}
	\end{tabular}%
	\caption{
		\textbf{Calculated magnetic moments as a function of doping.} The total magnetic moment of Tb(Mn$_{1-x}$Cr$_x$)$_6$Sn$_6$ as a function of $x$ is shown. It is well established that the interactions between in-plane Mn atoms are strongly FM \cite{riberolles2022prx}, but the nature of the interactions between Cr atoms remains unclear. Two cases are separately presented: one assuming only FM interactions between Cr atoms (Blue) and the other including AFM interactions between Cr atoms (Red). The black solid line represents the calculation under the assumption that, with Cr doping, the magnetic moments at specific atomic sites remain constant at 9.33, -2.42, -1.11, and 1.21 $\mu_B$ for Tb, Mn, Cr, and the sum of all other contributions, respectively. In the VC calculation (Green), all interactions between the 3$d$ transition metal atoms are assumed to be FM.}
	\label{fig:mm}
\end{figure}

~\rfig{fig:mm} presents the total magnetic moment (including the Tb orbital magnetic moment) of Tb(Mn$_{1-x}$Cr$_x$)$_6$Sn$_6$ as a function of $x$, which represents the Cr doping concentration. At $x = \frac{1}{6}$, the virtual crystal (VC) calculation and substitution calculation show similar trends, indicating that the rigid band model can explain the magnetic moment changes in this composition. In the rigid band model, an increasing magnetic moment with Cr substitution for Mn implies that the system loses more spin-down electrons than spin-up electrons as the Fermi energy (E$_F$) lowers. This behavior is related to the details of the density of states below E$_F$. For larger $x$, however, the results of the two methods deviate. While the VC calculation predicts an increasing total magnetic moment magnitude up to $x = 0.4$, the Cr atom substitution calculation shows a reduction in total magnetic moment after $x = \frac{1}{6}$. At this doping level, the rigid band model becomes unsuitable for describing the magnetic structure. The blue squares represent the total magnetic moment of the lowest energy configuration as a function of $x$. These calculations were performed under the hypothesis that the magnetic interaction between Cr atoms is ferromagnetic (FM), similar to the interaction between Mn atoms.

Next, since the magnetic interaction between Cr atoms in this compound need not have the same FM sign as in the host Mn plane, we extended the study to include antiferromagnetic (AFM) interactions between Cr atoms. This way, we found spin configurations with lower energy than the FM configuration (see ~\rtbl{tab:mm}, ~\rtbl{tab:mm1}), indicating the Cr atoms couple antiferromagnetically. In ~\rfig{fig:mm}, red asterisks represent the results of including AFM interactions. The most significant change occurs at $x = \frac{1}{3}$, where two substituted Cr atoms exhibit AFM interactions and cancel each other's magnetic moments. This results in a total magnetic moment of approximately $-0.3 \mu_B$ at $x = \frac{1}{3}$, compared to FM-only calculations, where the total magnetic moment reaches $-0.3 \mu_B$ only at $x = \frac{1}{2}$.

While the calculations show reasonable agreement with experiments regarding the magnetic moment compensation concentration (0.52 \textit{vs.} 0.43), the total magnetic moment change with $x$ does not align with experimental findings, which show a linear change with Cr concentration. In this calculation, we employed a unit cell for uniform doping, representing a highly ordered and limited scenario that may differ significantly from experimental conditions. Addressing this discrepancy would require using larger supercells to account for an doping disorder.

A simple method to estimate the compensation composition involves using a linear equation under the hypothesis that Cr doping does not alter the magnetic moments of individual atoms but merely substitutes Mn atoms with Cr atoms. The black solid line represents the calculation under this hypothesis. The magnetic moments of Tb, Mn, and the sum of all other atoms except Cr are taken from TbMn$_6$Sn$_6$, which are 9.33, $-2.42$, and 1.21 $\mu_B$, respectively. The Cr magnetic moment is obtained from TbCr$_6$Sn$_6$, which is $-1.11 \mu_B$. This calculation yields $x = 0.51$ for the compensation concentration, and the slope of the linear equation is 1.31 $\mu_B$/Cr, smaller than the experimental finding of 1.8 $\mu_B$/Cr.

~\rtbl{tab:mm} presents the averaged spin magnetic moment of each atom ($\mu_B$/atom), total magnetic moment ($\mu_B$/cell), and total energy (eV/cell) for various configurations and Cr concentrations. The Tb magnetic moment remains almost constant regardless of doping. The Mn atoms show variations in their averaged magnetic moments, though the changes are not significant. However, the Cr magnetic moment exhibits strong variations with structural configurations, demonstrating sensitivity to neighboring atoms. States with higher Cr average magnetic moments also have higher total energies, indicating that FM magnetic interactions between Cr atoms are energetically unfavorable.
\begin{table}[htb]
	\centering
	\bgroup
	\def\arraystretch{1.3}
	\begin{tabular*}{\linewidth}{c@{\extracolsep{\fill}}cccccc}		
			\toprule
			\textbf{x} & \textbf{config} & \textbf{total}&\textbf{Tb}&\textbf{Mn}&\textbf{Cr}& \textbf{$\Delta$ E (eV/cell)}\\ 
			\hline
			0  & 1 & -6.98& 6.33 &-2.42 &    &  \\ 
			\hline
			$\frac{1}{6}$  & 1 & -7.43& 6.30 &-2.55 &-1.92 &  \\ 
			\hline
			\multirow{4}{*}{$\frac{1}{3}$} & 1 & -4.92& 6.28 &-2.63 &-0.75 & 0.000 \\ 
			& 2 & -7.43& 6.29 &-2.68 &-1.89 & 0.121 \\ 
			& 3 & -7.49& 6.28 &-2.61 &-2.06 & 0.145 \\
			& 4$^*$ & -3.37& 6.26 &-2.60 & 0.11 &-0.046 \\
			\hline
			\multirow{4}{*}{$\frac{1}{2}$} & 1 & -3.23& 6.27 &-2.56 &-0.85 &  0.000 \\ 
			& 2 & -4.35& 6.27 &-2.78 &-0.99 & 0.013 \\
			& 3 & -4.74& 6.27 &-2.73 &-1.16 & 0.059 \\
			& 4$^*$ & -3.33& 6.26 &-2.69 &-0.71 &-0.075 \\
			
			\hline\hline
		\end{tabular*}
		\egroup
		\caption{
			Spin magnetic moments of Tb(Mn$_{1-x}$Cr$x$)$6$Sn$6$ are presented for various configurations. The total spin magnetic moment is given in units of $\mu_{B}$/cell. The magnetic moments of Mn and Cr atoms are averaged across all sites within the given configuration and are reported in units of $\mu_{B}$/atom. To account for the orbital contribution, 3.0 $\mu_{B}$ should be added for each Tb atom. Configuration 4$^*$ includes antiferromagnetic (AFM) interactions between Cr atoms. Total energies are compared between configurations with the same Cr concentrations.}
		\label{tab:mm}
	\end{table}

\begin{figure}[htb]
	\centering
	\begin{tabular}{c}
		\includegraphics[width=0.5\linewidth,clip]{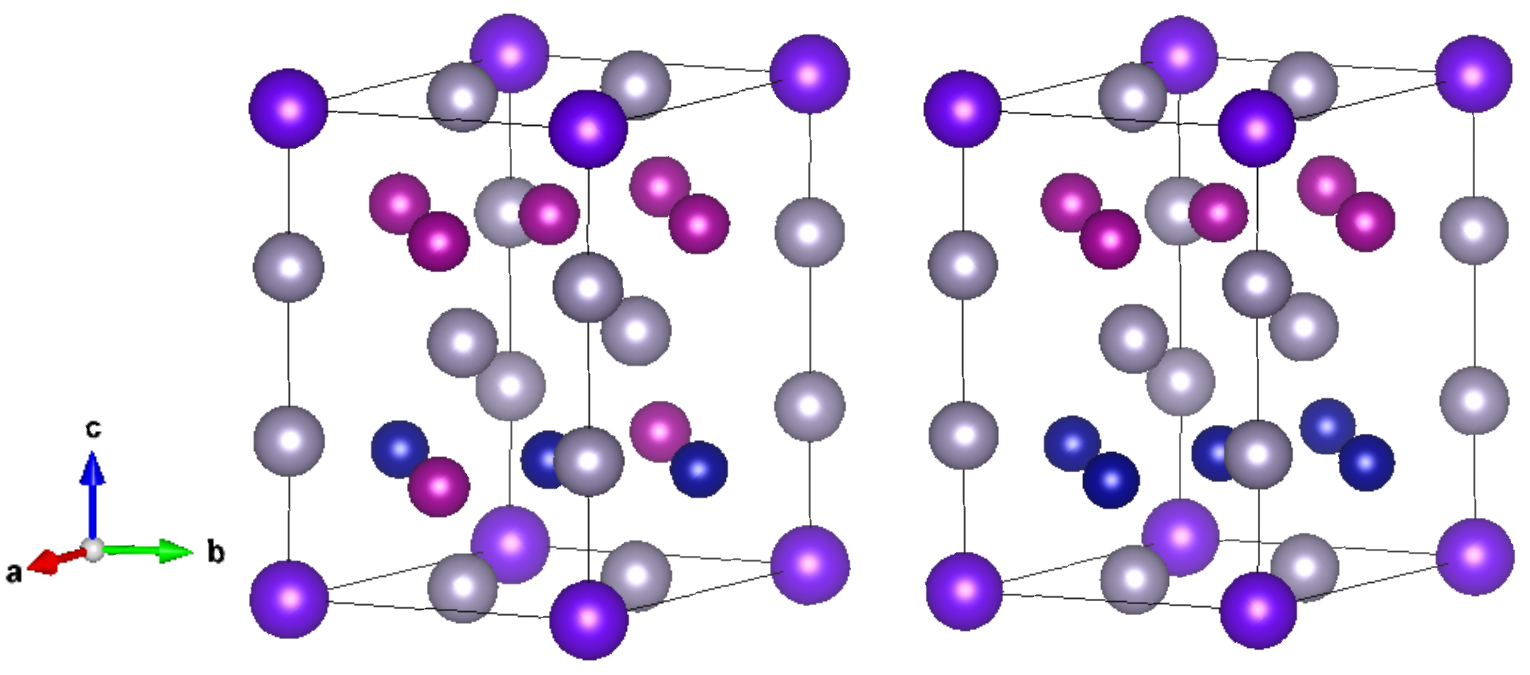}\\
		\includegraphics[width=0.5\linewidth,clip]{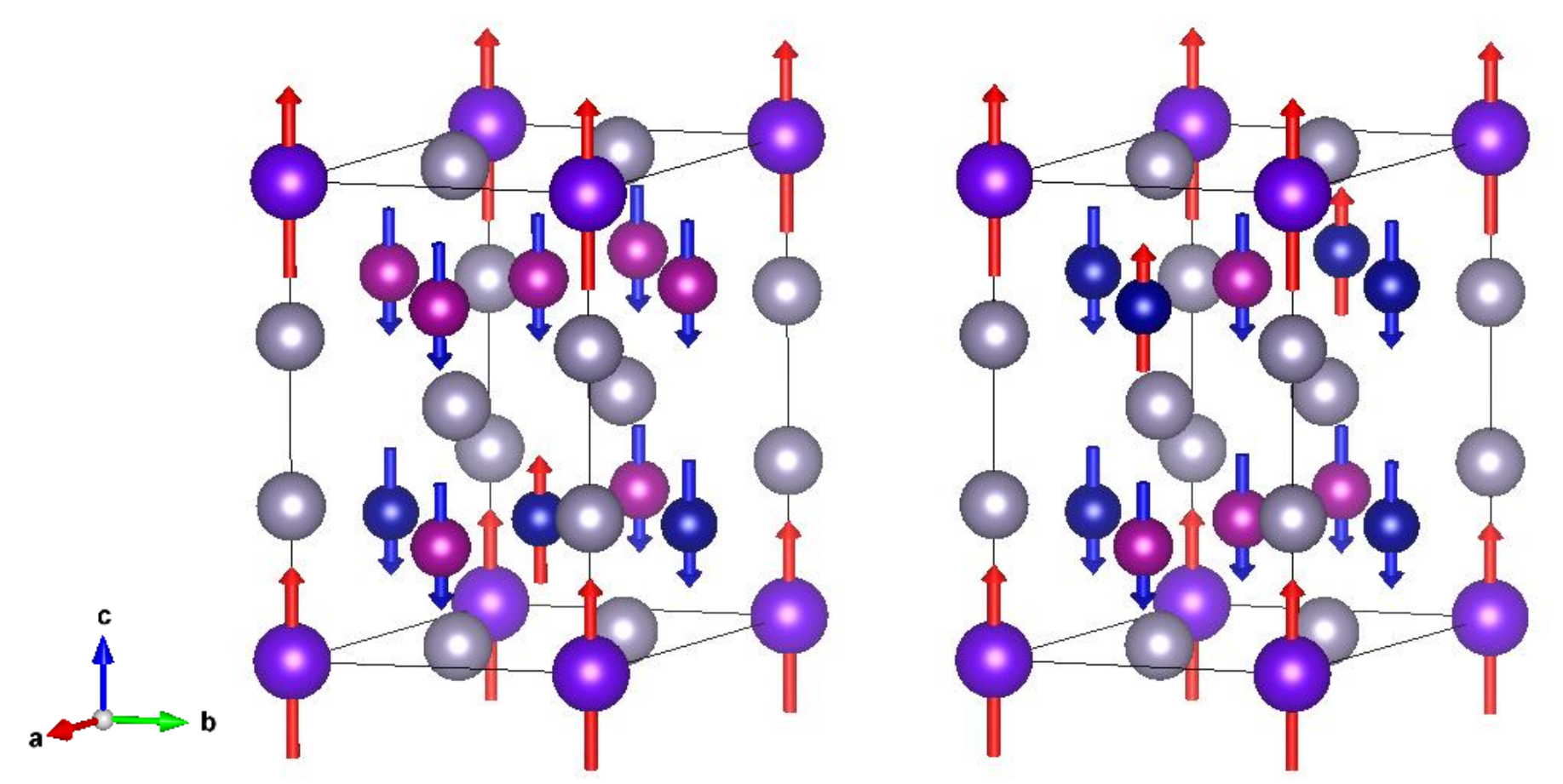}

	\end{tabular}%
	\caption{\textbf{Calculated magnetic configurations.} Top: The configurations of the lowest-energy states (config. 1) for $x=\frac{1}{3}$ (left) and $x=\frac{1}{2}$ (right), considering only FM interactions between transition metal atoms.\\
	Bottom: The spin configurations of the lowest-energy states (config. 4*) for  $x=\frac{1}{3}$ (left) and $x=\frac{1}{2}$  (right) are shown. In both cases, two Cr atoms within the same plane exhibit AFM interaction. The purple, red, blue, and gray spheres represent Tb, Mn, Cr, and Sn atoms, respectively. Red arrows indicate positive magnetic moments, while blue arrows represent negative magnetic moments.}
	\label{fig:struct}
\end{figure}

~\rtbl{tab:mm1} presents magnetic moments of each Mn and Cr atoms of the configurations that  have AFM interactions between some of Cr atoms. It displays the variation of magnetic moments which depend on the details of neighboring atoms.
The  Mn atom which has the largest magnetic moment is located on the same plane with two AFM Cr atoms in both $x=\frac{1}{3}$ and $x={1}{2}$. The Mn2 atom which has smallest magnetic moment among Mn atom in $x=\frac{1}{3}$ is located the top of Mn4 while other two (Mn1, Mn3) are the top of two Cr atom. ~\rfig{fig:struct} present the spin structure of these two lowest energy configuration.

~\rtbl{tab:mm1} presents the magnetic moments of individual Mn and Cr atoms in configurations where AFM interactions occur between certain Cr atoms. It shows the variation in magnetic moments, which depend on the details of neighboring atomic arrangements. The Mn atom with the largest magnetic moment in both $x = \frac{1}{3}$ and $x = \frac{1}{2}$ is located on the same plane as two AFM Cr atoms. Conversely, the Mn2 atom, which has the smallest magnetic moment among Mn atoms at $x = \frac{1}{3}$, is positioned directly above Mn4, while the other two Mn atoms (Mn1 and Mn3) are located above two Cr atoms. These two lowest-energy configurations are visualized in ~\rfig{fig:struct}.

\begin{table}[htb]
	\centering
	\bgroup
	\def\arraystretch{1.3}
	\begin{tabular*}{\linewidth}{c@{\extracolsep{\fill}}cccccccc}	
			\toprule
			\textbf{x} & \textbf{config} & \textbf{Mn1}&\textbf{Mn2}&\textbf{Mn3}&\textbf{Mn4}& \textbf{Cr1} & \textbf{Cr2} &\textbf{Cr3} \\ 
			\hline
			$\frac{1}{3}$  & 4$^*$ & -2.58& -2.49 &-2.61 & -2.74 & -2.10 & 2.12&  \\ 
			\hline
			$\frac{1}{2}$  & 4$^*$ & -2.65& -2.69 &-2.74 & &-2.18 & 2.08&-2.03 \\ 
			
			\hline\hline
	\end{tabular*}
	\egroup
	\caption{
		The spin magnetic moments of Mn and Cr atoms in the lowest-energy configuration are presented. For the $x = \frac{1}{3}$ ($\frac{1}{2}$) composition, there are 4 (3) Mn sites and 2 (3) Cr sites. The magnetic moments exhibit variations depending on the specific atomic sites.} 
	\label{tab:mm1}
\end{table}

\section{Supplementary Note 3: Temperature dependence of $\mu_0H_c$ near $x^*$}

\begin{figure*}
    \centering
    \includegraphics[width=0.9\textwidth]{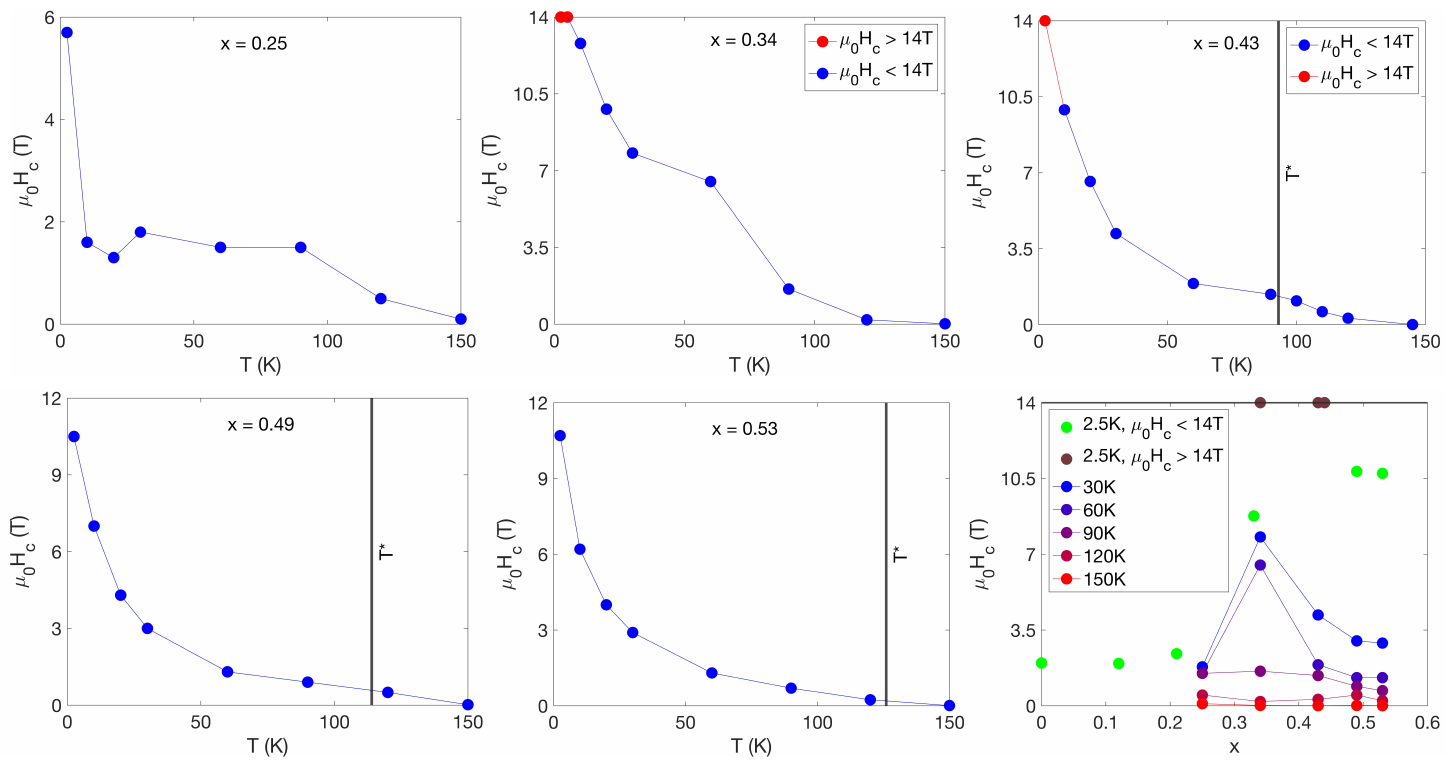}
    \caption{\textbf{Temperature dependence of $\mu_0H_c$ near $x^*$.} \textbf{a, b, c, d, e,} $\mu_0H_c$ as a function of temperature for several samples near $x^*$. Blue data points indicate $\mu_0H_c <$ \SI{14}{T} while red data points indicate $\mu_0H_c >$ \SI{14}{T}. \textbf{f,} $\mu_0H_c$ as a function of $x$ at a variety of temperatures.}
    \label{fig:tempdepofHc}
\end{figure*}

Fig.~\ref{fig:tempdepofHc} presents the full temperature dependence of $\mu_0H_c$ for samples near $x^*$. Panels a-e show $\mu_0H_c$ as a function of temperature for several samples with blue data points indicating $\mu_0H_c <$ \SI{14}{T} and red data points indicating $\mu_0H_c >$ \SI{14}{T}. Panel f presents these same data along with the full \SI{2}{K} $x$ dependence by plotting $\mu_0H_c$ as a function of $x$ at a variety of temperatures. It can be seen that $\mu_0H_c$ grows in all samples as the temperature decreases. There is no obvious change in $\mu_0H_c$ near $T^*$ as sometimes seen in compensated ferrimagnets~\cite{inhomoFeGdFiM, Mn2RuxGa_HcdivergingatTstar}.

\section{Supplementary Note 4: Analysis of AHE}

\begin{figure*}
    \centering
    \includegraphics[width=0.9\textwidth]{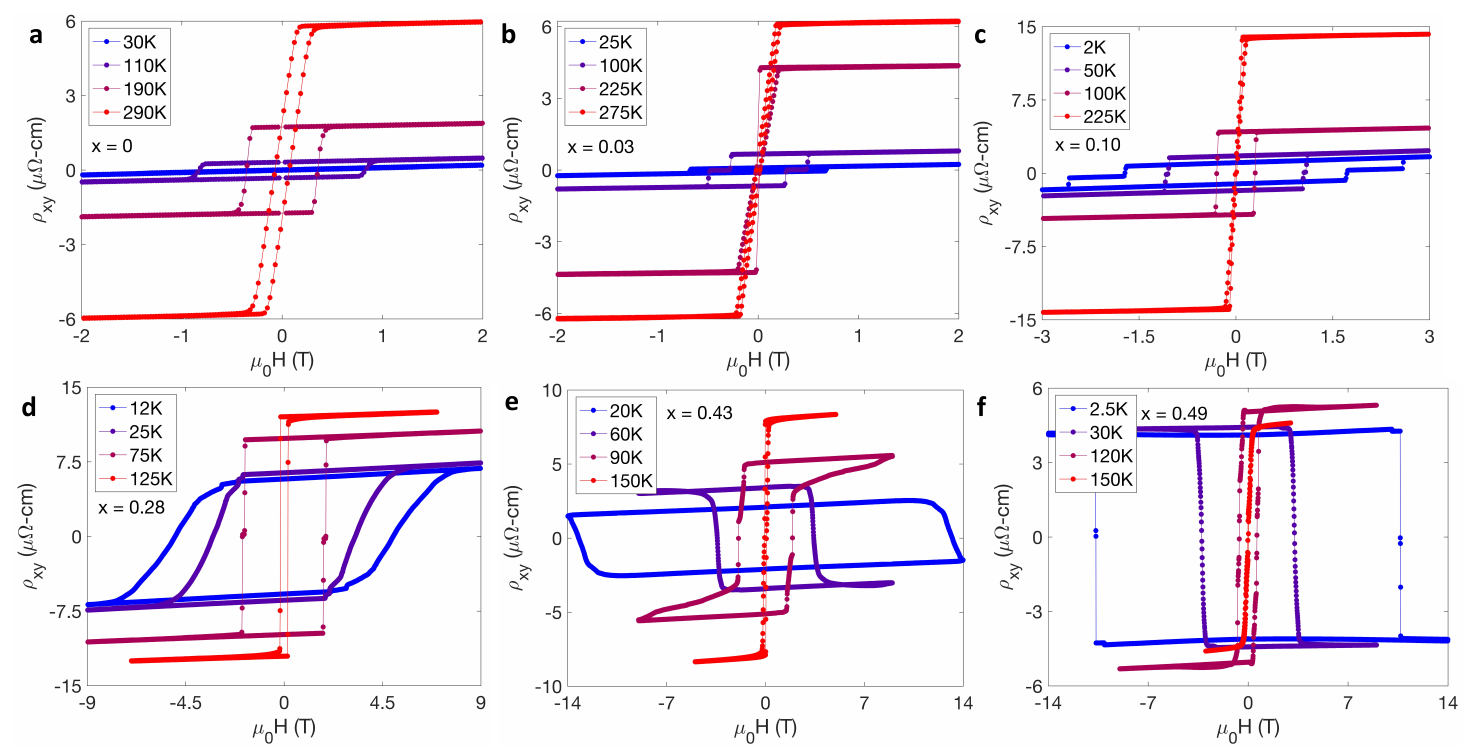}
    \caption{\textbf{Anomalous Hall effect of Tb(Mn$_{1-x}$Cr$_x$)$_6$Sn$_6$.} \textbf{a, b, c, d, e, f,} $\rho_{xy}$ as a function of $\mu_0H$ at several temperatures for $x = 0, 0.03, 0.10, 0.28, 0.43,$ and 0.49, respectively.}
    \label{fig:rhoxyvHdoping_extended}
\end{figure*}

Fig.~\ref{fig:rhoxyvHdoping_extended} presents $\rho_{xy}$ as a function of $\mu_0H$ at several temperatures for $x = 0, 0.03, 0.10, 0.28, 0.43,$ and 0.49, respectively. As noted in the main paper, a well-defined anomalous Hall effect can be observed.

\begin{figure}[h]
    \centering
    \includegraphics[width=0.5\textwidth]{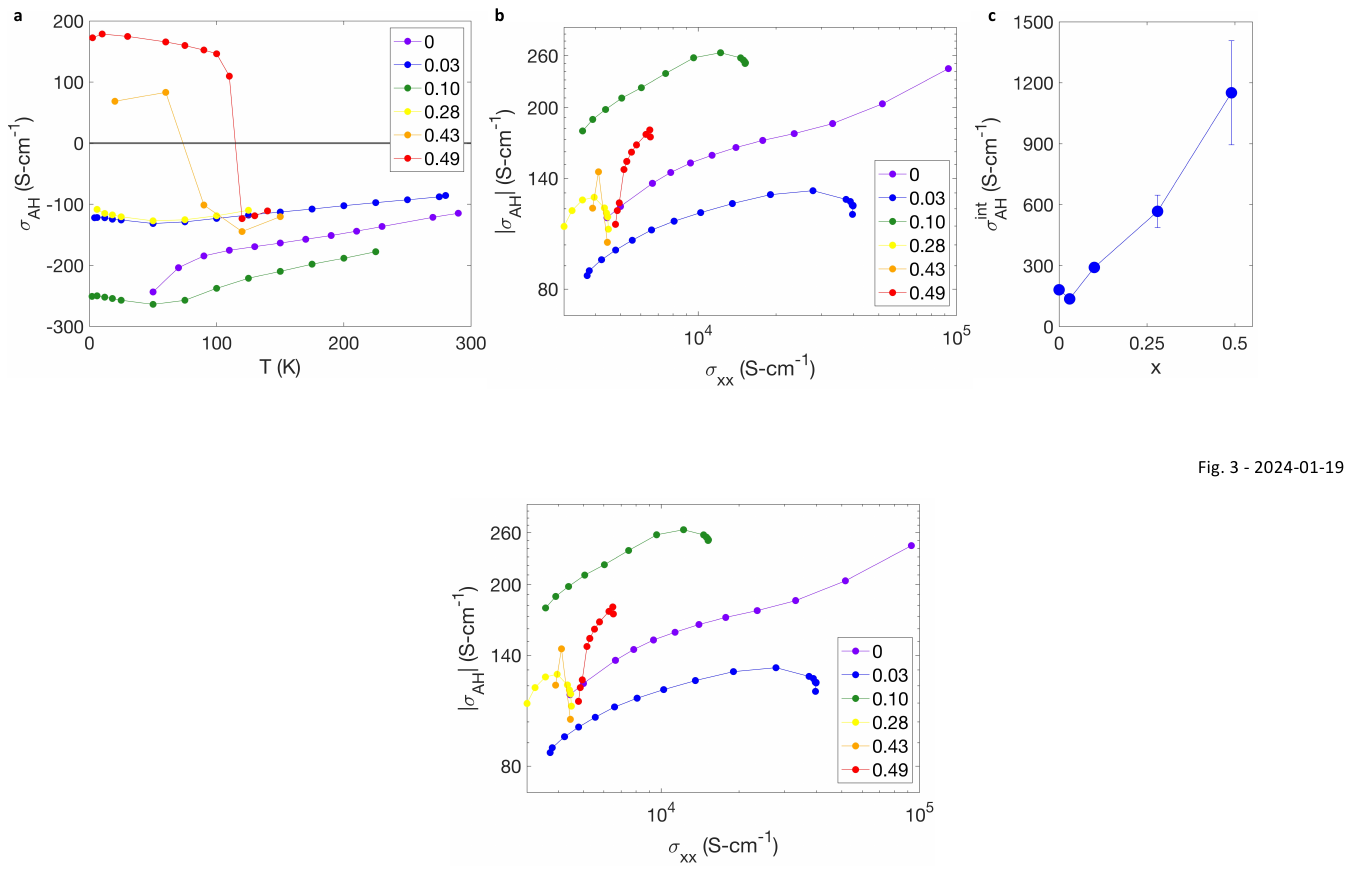}
    \caption{\textbf{Scaling of $|\sigma_{AH}|$ with $\sigma_{xx}$.} $|\sigma_{AH}|$ plotted as a function of $\sigma_{xx}$ on a log-log scale for various dopings of Tb(Mn$_{1-x}$Cr$_x$)$_6$Sn$_6$.}
    \label{fig:sigmaAHvsigmaxx}
\end{figure}

Fig.~\ref{fig:sigmaAHvsigmaxx} presents $|\sigma_{AH}|$ as a function of $\sigma_{xx}$ on a log-log scale for Tb(Mn$_{1-x}$Cr$_x$)$_6$Sn$_6$ for a variety of $x$. Here $\sigma_{xx}$ is defined as $\sigma_{xx} = \frac{\rho_{xx,0}}{\rho_{xx,0}^2+\rho_{AH}^2}$. For $x = 0.49$ data near $T^*$ are removed as $\rho_{AH}$ is very small as it crosses zero as a function of temperature and for $x = 0.43$ only data from $T > T^*$ are presented as there to a limited number of reliable data points available below $T^*$ due to the massive $H_c$ at low temperatures. These data imply that the AHE in Tb(Mn$_{1-x}$Cr$_x$)$_6$Sn$_6$ may be largely due to an intrinsic Berry phase contribution when they are compared to the theory from Ref.~\cite{AHEregimes}: regardless of $x$, $\sigma_{xx}$ is between 3$\times$10$^3$ and $10^5$ S-cm$^{-1}$ and $\sigma_{AH}$ is roughly constant (ie. does not change orders of magnitude) and of the order of 10$^2$ S-cm$^{-1}$ as $\sigma_{xx}$ is varied by over an order of magnitude.

\begin{figure*}[h]
    \centering
    \includegraphics[width=0.9\textwidth]{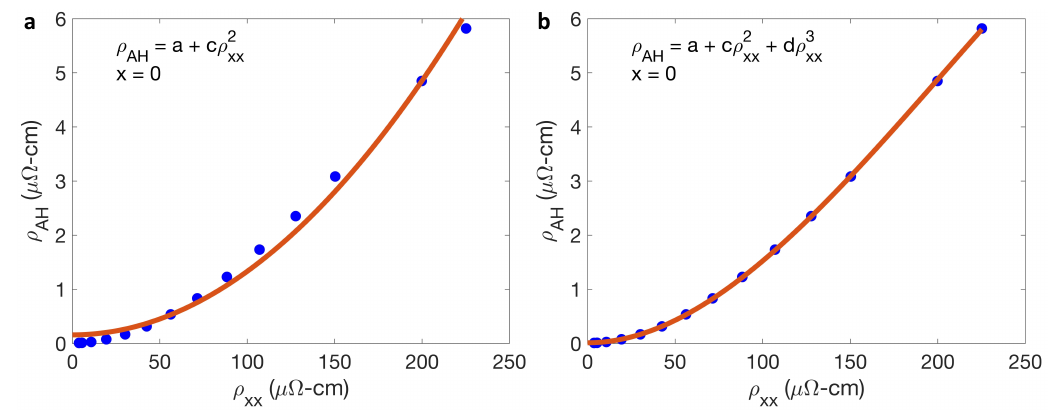}
    \caption{\textbf{Comparison of different AHE fitting functions.} \textbf{a,} Fitting of $\rho_{AH}$ vs. $\rho_{xx}$ for $x = 0$ using the standard fitting function of $\rho_{AH}=a+c\rho_{xx}^2$. \textbf{b,} Fitting of the same data to $\rho_{AH}=a+c\rho_{xx}^2+d\rho_{xx}^3$.}
    \label{fig:TMSAHEfittingcomp}
\end{figure*}

As noted in the main text, fitting of $\rho_{AH}$ as a function of $\rho_{xx}$ for \ch{TbMn6Sn6} is non-trivial. The resistivity form of the anomalous Hall fits are shown here to act as a check of the conductivity fits shwon in the main paper. The accepted scaling of the anomalous Hall effect in general goes as $\rho_{AH}=a+c\rho_{xx}^2$ where $c = \sigma_{AH}^{int}$~\cite{properscalingAHE}, yet adding an additional cubic term has been observed to improve the fit quality~\cite{originofSR_TMS}. This cubic term has been argued to be related to the fluctuations of Tb ions~\cite{originofSR_TMS} as noted in the main text. As seen in Fig.~\ref{fig:TMSAHEfittingcomp}, our data agree with this observation of a cubic term enhancing the fit quality in \ch{TbMn6Sn6} just as the $\frac{d}{\sigma_{xx}}$ term enhances the fit quality of the conductivity data. As such, we present analysis of the AHE using a fitting function of the form

\begin{equation}
    \rho_{AH}=a+c\rho_{xx}^2+d\rho_{xx}^3
    \label{eqn:AHEres}
\end{equation}

here.

\begin{figure*}
    \centering
    \includegraphics[width=0.9\textwidth]{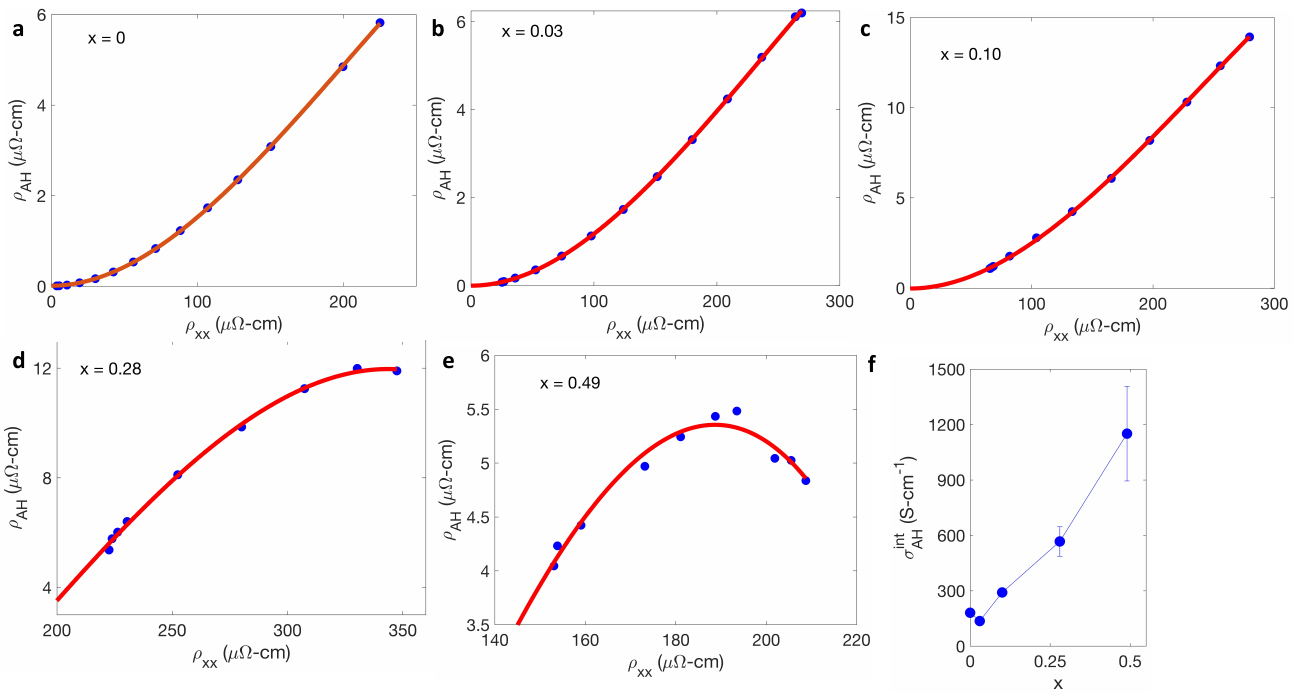}
    \caption{\textbf{AHE fitting using resistivity data.} \textbf{a, b, c, d, e,} Fitting of $\rho_{AH}$ vs. $\rho_{xx}$ for various $x$ using $\rho_{AH}=a+c\rho_{xx}^2+d\rho_{xx}^3$. \textbf{f,} Extracted $\sigma_{AH}^{int}$ as a function of $x$ from resistivity fittings in other panels.}
    \label{fig:rhoAHvrhoxxfits}
\end{figure*}

\begin{table}[h]
    \caption{Extracted parameters from resistivity fitting of AHE.}
    \centering
\begin{tabular}{||c | c | c | c | c||} 
 \hline
 $x$ & a ($\Omega$-cm) & c ([$\Omega$-cm]$^{-1}$) & d ([$\Omega$-cm]$^{-2}$) \\ [0.5ex] 
 \hline\hline
 0 & (1.27$\pm$0.94)$\times$10$^{-8}$ & 181$\pm$2.3 & (-2.96$\pm$0.17)$\times$10$^{5}$ \\ 
 \hline
 0.03 & (-4.11$\pm$14.7)$\times$10$^{-9}$ & 136$\pm$2.5 & (-1.86$\pm$0.10)$\times$10$^{5}$ \\
 \hline
 0.10 & (-2.19$\pm$7.21)$\times$10$^{-8}$ & 291$\pm$9.9 & (-4.01$\pm$0.35)$\times$10$^{5}$  \\
 \hline
 0.28 & (-1.04$\pm$0.20)$\times$10$^{-5}$ & 567$\pm$80 & (-1.10$\pm$0.19)$\times$10$^{6}$   \\ 
 \hline
 0.49 & (-8.31$\pm$2.72)$\times$10$^{-6}$ & 1151$\pm$256 & (-4.07$\pm$0.94)$\times$10$^{6}$  \\ [1ex] 
 \hline
\end{tabular}
    \label{tab:AHEresparameters}
\end{table}
\begin{center}
\end{center}

Fig.~\ref{fig:rhoAHvrhoxxfits}a-e present fits of $\rho_{AH}$ vs. $\rho_{xx}$ for a variety of $x$ using Eqn.~\ref{eqn:AHEres}. Note that for $T < T^*$ the sign of $\rho_{AH}$ has been changed to match the sign of data taken at $T > T^*$, data near $T^*$ have been removed as $\rho_{AH}$ becomes small as it changes sign, and that the lack of analysis for $x \approx x^*$ is due to the high $\mu_0H_c$ observed at low temperatures which limits the amount of data points that can be reliably fit to. Table~\ref{tab:AHEresparameters} lists the fitting parameters for these fits as well as their 95\% confidence bounds. Fig.~\ref{fig:rhoAHvrhoxxfits}f plots $\sigma_{AH}^{int}$ (or c) as a function of $x$. Note that these extracted parameters are very similar to those shown in the main paper using fitting to the conductivity.





\begin{figure*}
    \centering
    \includegraphics[width=0.9\textwidth]{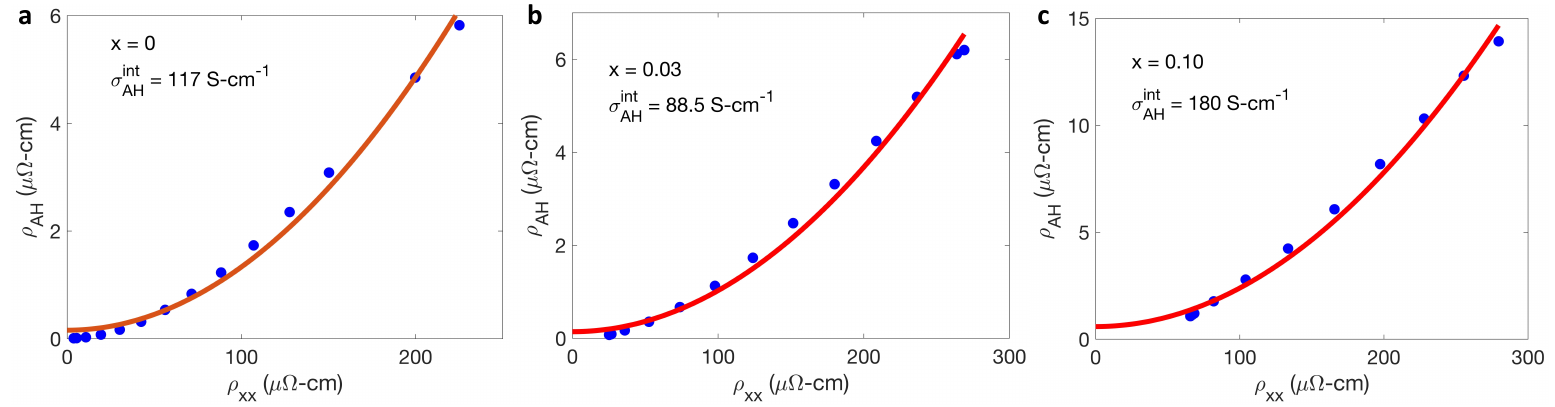}
    \caption{\textbf{AHE fitting using resistivity data without cubic term.} \textbf{a, b, c,} Fitting of $\rho_{AH}$ vs. $\rho_{xx}$ for low $x$ using $\rho_{AH}=a+c\rho_{xx}^2$.}
    \label{fig:rhoAHvrhoxxfits_nocubicterm}
\end{figure*}


As another check of the results presented here, Fig.~\ref{fig:rhoAHvrhoxxfits_nocubicterm} presents fitting of $\rho_{AH}$ vs. $\rho_{xx}$ for low $x$ using the standard AHE resistivity equation $\rho_{AH}=a+c\rho_{xx}^2$. Only low $x$ data are shown as the data in this regime is still quasi-quadratic. On the other hand, it can clearly be seen in Fig.~\ref{fig:rhoAHvrhoxxfits}d,e that for high $x$ the data becomes clearly not quadratic. While these fits reduce the extrapolated $\sigma_{AH}^{int}$ as compared to the the fits with the cubic terms, the results are qualitatively consistent: $x = 0.03$ has a smaller $\sigma_{AH}^{int}$ than $x = 0$ and $x = 0.10$ has a larger $\sigma_{AH}^{int}$ than both $x = 0$ and $x = 0.03$.

\section{Supplementary Note 5: Further Crystal Growth Details}

\begin{figure}[h]
    \centering
    \includegraphics[width=0.5\textwidth]{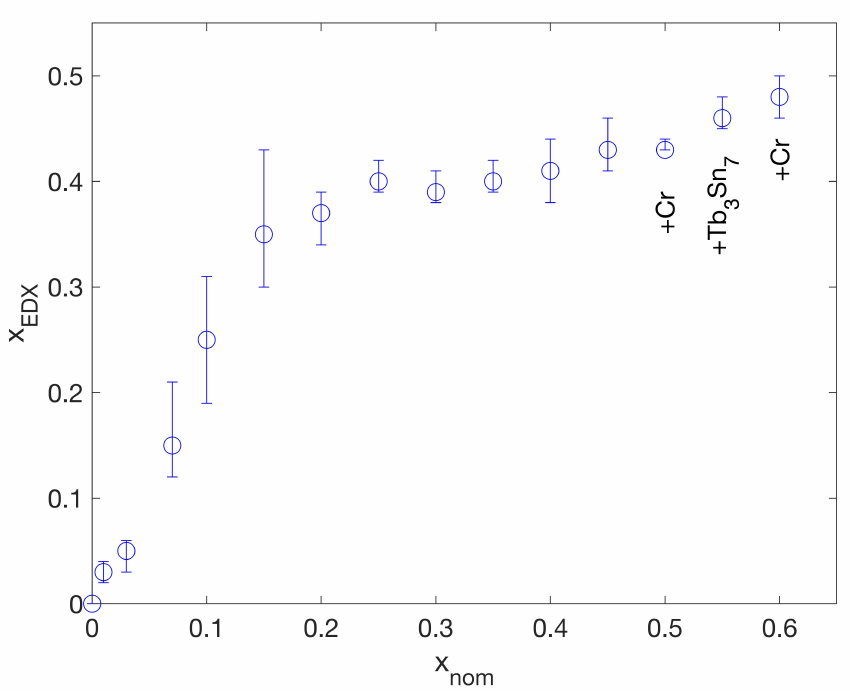}
    \caption{\textbf{$x_{EDX}$ as a function for $x_{nom}$ for a set of Tb(Mn$_{1-x}$Cr$_x$)$_6$Sn$_6$ growths.} The open circles indicate the average of $x_{EDX}$ found in the growth and the error bars indicate the range of $x_{EDX}$ in the growth. ``$+y$" indicates that a secondary phase $y$ was also found in the growth.}
    \label{fig:xedxxnom}
\end{figure}

\begin{figure*}
    \centering
    \includegraphics[width=0.9\textwidth]{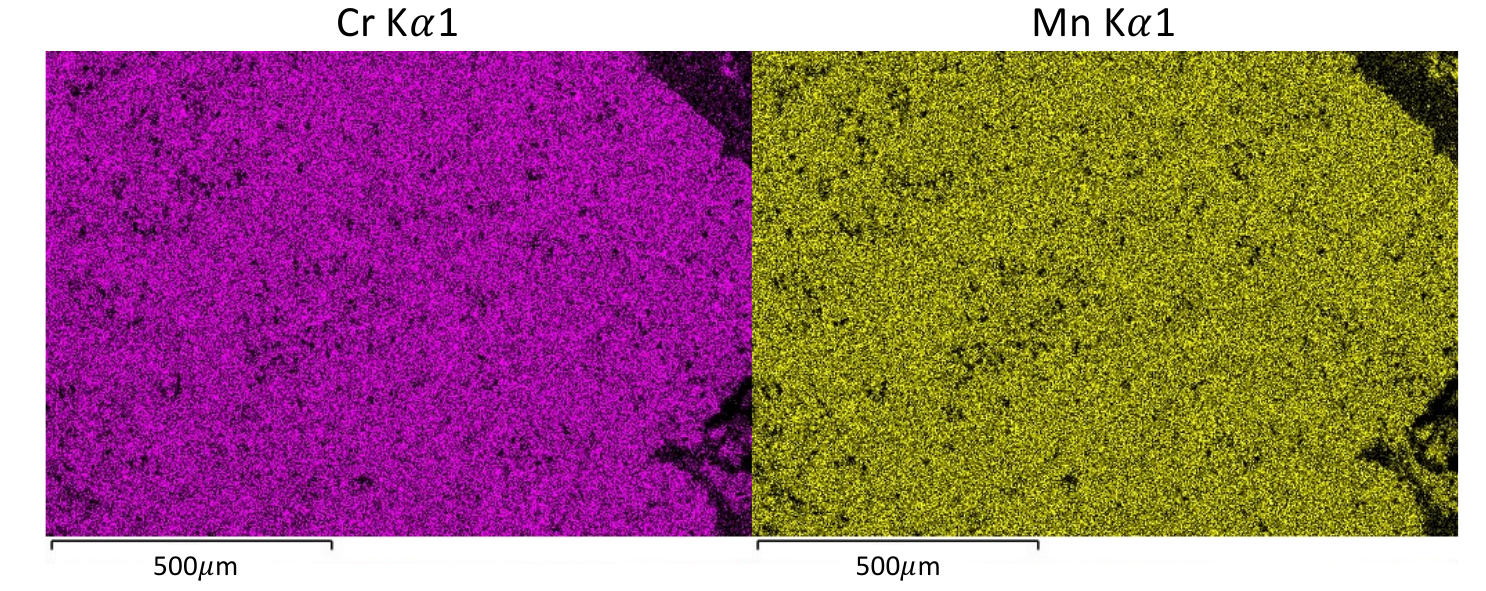}
    \caption{\textbf{EDX map of $x = 0.42$.} An EDX map of the Cr K$\alpha$1 (left) and Mn K$\alpha$1 (right) signal on a sample of Tb(Mn$_{1-x}$Cr$_x$)$_6$Sn$_6$ with $x = 0.42$. The scale bars are \SI{500}{\micro\metre}.}
    \label{fig:EDXmap}
\end{figure*}

Fig.~\ref{fig:xedxxnom} presents $x_{EDX}$ as a function for $x_{nom}$ for a set of Tb(Mn$_{1-x}$Cr$_x$)$_6$Sn$_6$ growths. The open circles indicate the average of $x_{EDX}$ found in the growth and the error bars indicate the range of $x_{EDX}$ in the growth. Each growth had between 3 and 8 crystals of Tb(Mn$_{1-x}$Cr$_x$)$_6$Sn$_6$ measured with at least 8 points measured on each crystal. ``$+y$" indicates that a secondary phase $y$ was also found in the growth. Cr and \ch{Tb3Sn7} crystals were identified in the growths with $x_{nom}\geq$ 0.5. The large spread in $x_{EDX}$ measured for each growth in the low $x_{nom}$ growths likely is related to the steepness of the $x_{EDX}$ vs. $x_{nom}$ curve in this regime. As mentioned in the main paper, growths with $x_{nom} >$ 0.6 using the growth recipe utilized here did not provide any crystals, likely due to the reduced solubility of Cr in Sn~\cite{CrSn_binaryphasediag} as compared to the solubility of Mn in Sn~\cite{MnSn_binaryphasediag}. This is also evidenced by the precipitation of Cr crystals from the melt at high $x_{nom}$. The EDX statistics for several dopings are given in the tables below, and indicate a reasonably homogeneous doping across the entire doping range. Furthermore, an EDX map for a sample with $x = 0.42$ is shown in Fig.~\ref{fig:EDXmap} with the left panel showing the Cr K$\alpha$1 response and the right panel showing the Mn K$\alpha$1 response. $x$ was determined with point and ID mode of the EDX system as it is more quantitatively accurate, but mapping mode offers further evidence that the doping is qualitatively consistent throughout a crystal. Note that the scale bars are \SI{500}{\micro\metre}, thus this map is done over a region that is larger than the area of a typical sample used for the magnetization and magnetotransport measurements.

\begin{table}[h]
    \caption{Supplementary Table 3. EDX statistics for $x = 0.04$ in atomic percents.}
    \centering
\begin{tabular}{||c | c | c | c | c||} 
 \hline
  & Tb & Cr & Mn & Sn \\ [0.5ex] 
 \hline\hline
 Maximum & 7.83 & 3.00 & 47.13 & 44.89 \\ 
 \hline
 Minimum & 7.31 & 1.82 & 45.66 & 43.51 \\
 \hline
 Average & 7.55 & 2.12 & 46.13 & 44.21 \\
 \hline
 Standard Deviation & 0.16 & 0.40 & 0.50 & 0.50 \\
 [1ex] 
 \hline
\end{tabular}
    \label{tab:EDXforp04}
\end{table}
\begin{center}
\end{center}

\begin{table}[h]
    \caption{Supplementary Table 4. EDX statistics for $x = 0.35$ in atomic percents.}
    \centering
\begin{tabular}{||c | c | c | c | c||} 
 \hline
  & Tb & Cr & Mn & Sn \\ [0.5ex] 
 \hline\hline
 Maximum & 7.78 & 17.63 & 32.28 & 45.13 \\ 
 \hline
 Minimum & 7.14 & 16.76 & 30.53 & 42.66 \\
 \hline
 Average & 7.48 & 17.11 & 31.49 & 43.93 \\
 \hline
 Standard Deviation & 0.19 & 0.34 & 0.76 & 0.79 \\
 [1ex] 
 \hline
\end{tabular}
    \label{tab:EDXforp35}
\end{table}
\begin{center}
\end{center}

\begin{table}[h]
    \caption{Supplementary Table 5. EDX statistics for $x = 0.50$ in atomic percents.}
    \centering
\begin{tabular}{||c | c | c | c | c||} 
 \hline
  & Tb & Cr & Mn & Sn \\ [0.5ex] 
 \hline\hline
 Maximum & 7.82 & 24.46 & 24.94 & 44.71 \\ 
 \hline
 Minimum & 7.23 & 23.40 & 23.64 & 43.24 \\
 \hline
 Average & 7.57 & 24.13 & 24.45 & 43.85 \\
 \hline
 Standard Deviation & 0.21 & 0.39 & 0.43 & 0.21 \\
 [1ex] 
 \hline
\end{tabular}
    \label{tab:EDXforp50}
\end{table}
\begin{center}
\end{center}

\section{Supplementary Note 6: Two-State Magnetic Writing and Reading}

\begin{figure*}
    \centering
    \includegraphics[width=0.9\textwidth]{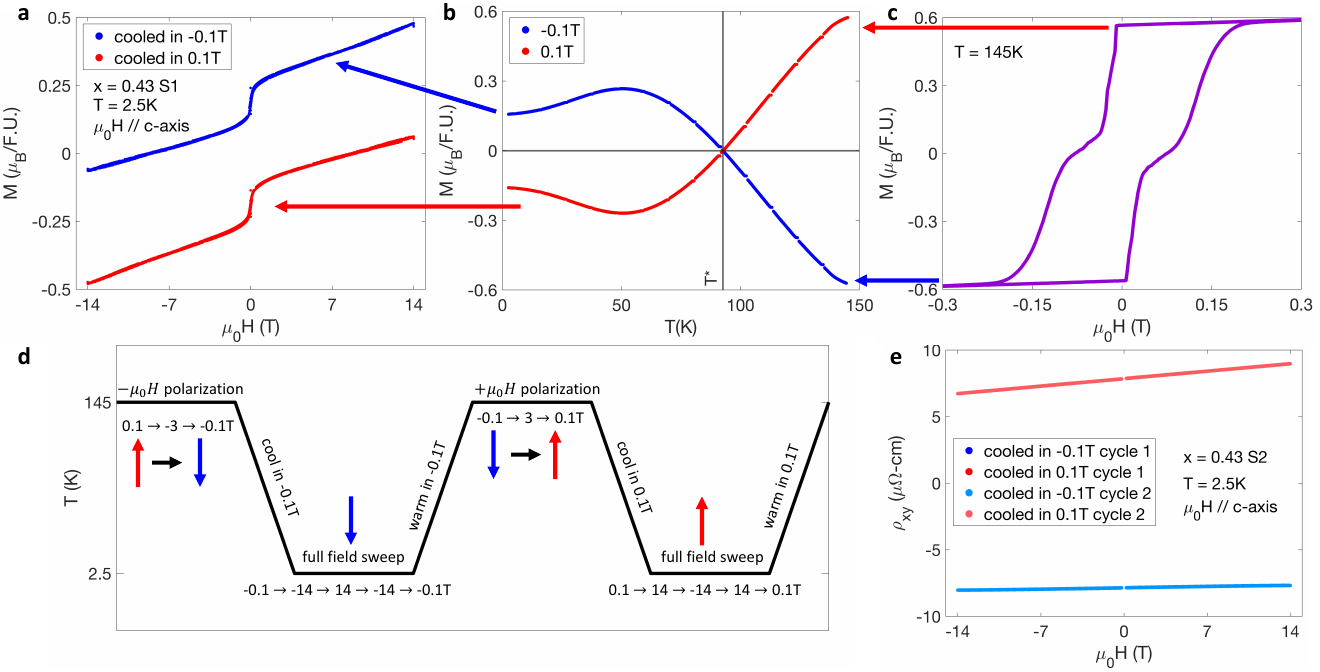}
    \caption{\textbf{Two-state magnetic writing and reading.} \textbf{a,} Magnetization as a function of magnetic field at \SI{2.5}{K} for $x = 0.43$ after being trained and subsequently cooled in small positive and negative magnetic fields. \textbf{b,} Magnetization as a function of temperature while cooling in small magnetic fields after training. \textbf{c,} Full magnetization versus magnetic field hysteresis loop measured at \SI{145}{K}. Note that this sample is a hard ferromagnet with a small coercive field at this temperature. \textbf{d,} Sequence that can be used to utilize Tb(Mn$_{1-x}$Cr$_x$)$_6$Sn$_6$ near $x^*$ as a simple memory device by virtue of the temperature dependence of $\mu_0H_c$. The colored arrows indicate the direction of the net magnetic moment. At high temperatures $\mu_0H_c$ is small, allowing for polarization into a chosen state with a relatively small field. At low temperatures $\mu_0H_c$ is larger than \SI{14}{T}, effectively ``locking in'' the state that was chosen at high temperatures. \textbf{e,} $\rho_{xy}$ for another sample of $x = 0.43$ as a function of magnetic field at \SI{2.5}{K} after cooling in a small positive or negative field. Note that the curves of cycles 1 and 2 fall on top of each other at this resolution.}
    \label{fig:memory}
\end{figure*}

As seen elsewhere in this paper, in Tb(Mn$_{1-x}$Cr$_x$)$_6$Sn$_6$ $\mu_0H_c$ grows substantially as temperature decreases, especially in samples near $x^*$. Fig.~\ref{fig:memory} focuses on $x = 0.43$ which is close to the experimentally observed $x^*$ and exhibits $T_{sr} =$ \SI{148}{K}. Fig.~\ref{fig:memory}a presents the magnetization as a function of magnetic field applied along the c-axis after being cooled from high temperatures in a magnetic field of 0.1 or \SI{-0.1}{T}. These two curves are part of a hysteresis loop with $H_c >$ \SI{14}{T}, and thus they have different values across the whole measured field range. Panels b and c show the process to reach these two distinct low temperature states: at \SI{145}{K} the sample is a hard ferromagnet with a relatively small coercive field. Due to this, the sample can be polarized with relatively small magnetic fields. After the sample has been polarized at this high temperature and subsequently cooled in either a small negative or positive field to \SI{2.5}{K}, the moments will be ``locked in'' and unable to be changed to the other state. It should be noted that this process also works without applying a small negative field while cooling (ie. if a sample is trained with a positive magnetic field at \SI{145}{K}, then the magnetic field is turned completely off, then the sample is cooled, the same low temperature results will be obtained).

Using this same principle of a highly temperature dependent coercive field, a simple memory device can be made utilizing the AHE in Tb(Mn$_{1-x}$Cr$_x$)$_6$Sn$_6$. Fig.~\ref{fig:memory}d presents a process that be used to accomplish task with the colored arrows indicating the direction of the net magnetic moment: the moments are polarized at \SI{145}{K} with a relatively small magnetic field, then cooled in a small magnetic field of the same sign (or zero magnetic field as described above), then a full magnetic field sweep up to fields of $\pm$\SI{14}{T} is performed at \SI{2.5}{K}. After warming back to \SI{145}{K}, the process can be repeated with magnetic fields of the opposite signs. This entire process constitutes one ``cycle''. Panel e shows the Hall effect measured at \SI{2.5}{K} for another sample of $x = 0.43$ from two cycles. Note that the data from cycle 1 and cycle 2 overlap each other on this scale. Due to the relatively large $\rho_{AH}$ and extremely large $\mu_0H_c$ in samples near $x^*$, a large difference in the $\rho_{xy}$ can be observed between data obtained after training in positive or negative magnetic fields at high temperatures. This offers a proof of concept of the enhanced AHE in Tb(Mn$_{1-x}$Cr$_x$)$_6$Sn$_6$ being utilized as a simple way to easily write a magnetic state at relatively high temperatures and subsequently read the state at low temperatures with the state remaining robust up to at least \SI{14}{T}. 

\clearpage

\bibliography{supp}